# Advancing the Pareto front for thin-film materials using a self-driving laboratory

Benjamin P. MacLeod[1,2,*], Fraser G. L. Parlane[1,2,*], Connor C. Rupnow[1,2,3], Kevan E. Dettelbach[1], Michael S. Elliott[1], Thomas D. Morrissey[1,2], Ted H. Haley[1], Oleksii Proskurin[1], Michael B. Rooney[1], Nina Taherimakhsousi[1], David J. Dvorak[2], Hsi N. Chiu[1], Christopher E. B. Waizenegger[1], Karry Ocean[1], Mehrdad Mokhtari[1] & Curtis P. Berlinguette[1,2,3,4,†]

[1]Department of Chemistry, The University of British Columbia, 2036 Main Mall, Vancouver, BC V6T 1Z1, Canada.
[2]Stewart Blusson Quantum Matter Institute, The University of British Columbia, 2355 East Mall, Vancouver, BC V6T 1Z4, Canada.
[3]Department of Chemical and Biological Engineering, The University of British Columbia, 2360 East Mall, Vancouver, BC V6T 1Z3, Canada.
[4]Canadian Institute for Advanced Research (CIFAR), MaRS Centre, 661 University Avenue Suite 505, Toronto, ON M5G 1M1, Canada.

*These authors contributed equally to this work.
†Corresponding author. Email: cberling@chem.ubc.ca

**Abstract**

Useful materials must satisfy multiple objectives, where the optimization of one objective is often at the expense of another. The Pareto front reports the optimal trade-offs between competing objectives. Here we report a self-driving laboratory, "Ada", that defines the Pareto front of conductivities and processing temperatures for palladium films formed by combustion synthesis. Ada identified previously untested combustion synthesis conditions that resulted in the discovery of lower processing temperatures (below 200 °C) relative to the prior art for this technique (250 °C), a temperature difference that makes the coating of different commodity plastic materials possible (e.g., Nafion, polyethersulfone). These conditions enabled us to use combustion synthesis to spray coat uniform palladium films with moderate conductivity ($1.1 \times 10^5$ S m$^{-1}$) at 191 °C. Spray coating at 226 °C yielded films with conductivities ($2.0 \times 10^6$ S m$^{-1}$) comparable to those of sputtered films (2.0 to $5.8 \times 10^6$ S m$^{-1}$). This work shows how self-driving laboratories can discover materials satisfying multiple objectives.

**Introduction**

Self-driving laboratories combine automation and artificial intelligence to accelerate the discovery and optimization of materials[1–3]. The increasing flexibility of laboratory automation is enabling self-driving laboratories to manipulate and measure a broader set of experimental variables. Consequently, a growing number of self-driving laboratories are being developed across a range of fields[4–27]. While many self-driving laboratories are able to test multiple experimental variables, most optimize for a single objective (e.g., process parameter, material property)[6–15]. This situation is not consistent with most practical applications, where multiple objectives need to be simultaneously optimized[28–34]. Consider, for example, how a solar cell must be optimized for voltage, current, and fill factor to yield a high power conversion efficiency[35,36], how an electrolyzer must form products at low voltages and high reaction rates and selectivities[37], and how structural alloys are optimized for both strength and toughness[30,34]. These and other applications motivate the emerging use of self-driving laboratories for multiobjective optimization[17–27].

The optimization of materials for multiple objectives can be challenging because improving one objective often compromises another (e.g., decreasing the bandgap of the light-absorbing material in a photovoltaic cell increases the photocurrent but decreases the voltage[38]). As a result, there is often no single champion material, but rather a set of materials exhibiting trade-offs between objectives (Fig. 1). The set of materials exhibiting the best possible trade-offs lie at the Pareto front. Materials on the Pareto front cannot be improved for one objective without negatively impacting one or more other objectives. Most self-driving laboratories used for multiobjective optimization identify a single optimal material based on preferences specified in advance of the experiment[17–24]. Here we configure a self-driving laboratory to map out the *entire* Pareto front[26,27] and apply this approach to thin film materials for the first time. This approach is highly relevant to the materials sciences because it identifies optimal materials for every preferred tradeoff between objectives.

Building on our self-driving laboratory, "Ada"[7] (Fig. 2), we reconfigured the hardware and software to optimize for an entire Pareto front for the combustion synthesis of conducting palladium films. The self-driving laboratory was designed to map out a Pareto front that shows the tradeoff between the temperature at which the films are processed, and conductivity. We selected combustion synthesis as an optimization problem because it is a solution-based method for making functional metal coatings. This method, however, has not yet been scaled and has not been proven for making high-quality, conductive metal films[39–41]. Combustion synthesis can form coatings at lower temperatures, enabling the potential use of inexpensive polymeric substrates,[42,43] but film conductivity typically decreases with processing temperature[39]. This situation presents a trade-off: to what extent can the conductivity be maximized while the processing temperature is minimized? The answer to this question would enable the researcher to determine, for example, what types of substrates could be layered with a metal coating of certain conductivity. We therefore leveraged Ada to effectively study the numerous compositional[44,45] and processing variables[39,46,47] that influence processing temperatures and the corresponding conductivities. In doing so, we identified previously untested conditions that decreased the temperature required for the combustion synthesis of palladium from 250 °C to 190 °C. This finding increases the scope of polymeric substrates that palladium can be deposited on by combustion synthesis to include Nafion[48], polyethersulfone[49], and heat-stabilized polyethylene napththalate[49]. We were able to use these conditions to synthesize homogeneous films on larger substrates with conductivities that approach those of films made by vacuum deposition methods.

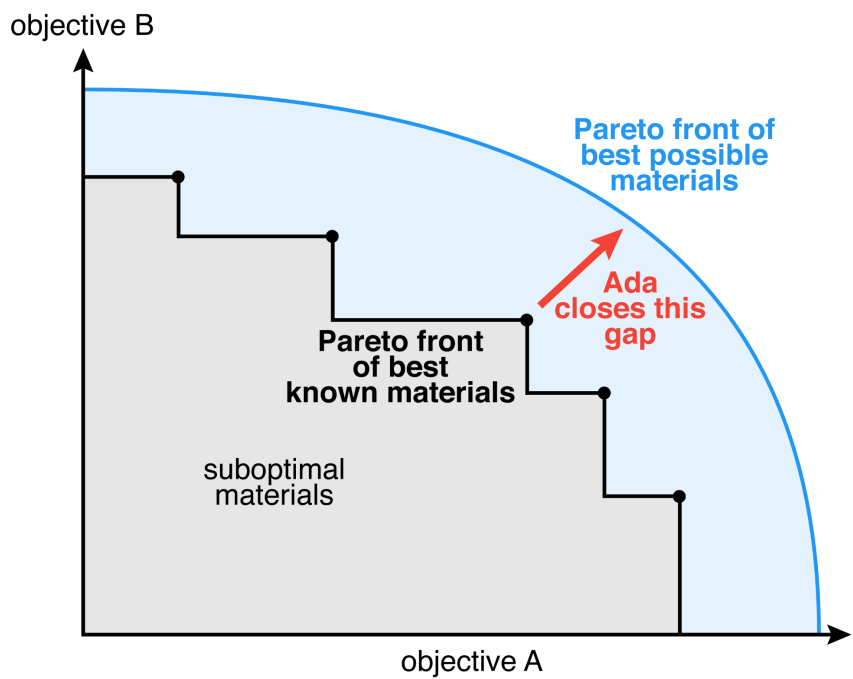

**Fig. 1 | A Pareto front.** No single optimal material exists when searching for materials that satisfy two or more conflicting objectives (e.g. film conductivity and processing temperature). Rather, an ideal set of possible materials exist with optimal tradeoffs between the objectives

(indicated by the blue curve). The state-of-the-art materials that offer the best compromises between the two objectives form the experimentally observed Pareto front (black points).

**Autonomously discovering a Pareto front**

For this study we configured Ada to manipulate four variables: fuel identity; fuel-to-oxidizer ratio; precursor solution concentration; and annealing temperature (Fig. 3a). We confined the study to mixtures of two fuels, glycine and acetylacetone, that we independently identified to yield conductive films at temperatures below 300 °C. The fuel-to-oxidizer ratio was varied because it controls product oxidation in bulk combustion syntheses[50,51]. The precursor concentration influences the morphology of the drop-casted films. Finally, we varied the processing temperature which may influence the conductivity through solvent removal, precursor decomposition, film densification, impurity removal, grain growth, oxidation, or cracking[52–54].

Flexible automation enabled us to construct Ada (Fig. 2a), which consists of a smaller, 4-axis robot coupled to a larger, 6-axis robot. The smaller robot (Fig. 2b) deposited and characterized the thin films[7], while the larger robot transported the samples to a commercial X-ray fluorescence (XRF) microscope for elemental analysis. These two robots jointly executed a 7-step experimental workflow (Fig. 2c, methods). First, a combustion synthesis precursor solution was formulated from stock solutions and then drop-cast onto a glass microscope slide. The resulting precursor droplet was imaged and then annealed in a forced-convection oven to form a film. The film was subsequently characterized by XRF microscopy, imaging, and 4-point probe conductance mapping. The conductivity of each film was determined indirectly by combining the conductance with a film thickness estimated by XRF (see autonomous workflow step 7 in methods, Supplementary Fig. 2). Finally, the conductivity and processing temperature for each film were passed to the qEHVI multiobjective Bayesian optimization algorithm[55] to plan the next experiment based on all the available data (see autonomous workflow step 8 in methods).

All of the steps in the autonomous workflow were performed without human intervention at a typical rate of 2 samples an hour. Ada could run unattended for 40-60 experiments until the necessary consumables (e.g., pipettes tips, mixing vials, glass substrates, and precursors; see methods) were exhausted. We used Ada to execute a total of 253 combustion synthesis experiments that explored a wide range of pertinent composition and processing variables.

The qEHVI algorithm is one of a number of *a posteriori* multiobjective optimization algorithms designed to identify the Pareto front[55–59]. These multiobjective optimization methods are known as *a posteriori* methods because preferred solutions are selected *after* the optimization. We chose to use an *a posteriori* method for this exploratory study because we sought to identify a range of Pareto-optimal outcomes rather than a single optimal point. We selected the qEHVI algorithm because previously reported benchmarks show that the qEHVI algorithm often resolves the Pareto front in fewer experiments than other algorithms.[55]

The qEHVI algorithm directed our self-driving laboratory to quantify the trade-off between film conductivity and annealing temperature (Fig. 3). We manually selected eight synthesis conditions spanning most of the design space to provide initialization data for the qEHVI algorithm (Supplementary Table 1). After executing these initial experiments, Ada executed more than 50 iterative qEHVI-guided experiments to map the Pareto front of annealing temperature and conductivity. We performed this autonomous optimization campaign in quadruplicate. Each replicate generated a Pareto front showing a clear trade-off between temperature and conductivity (Fig. 3b).

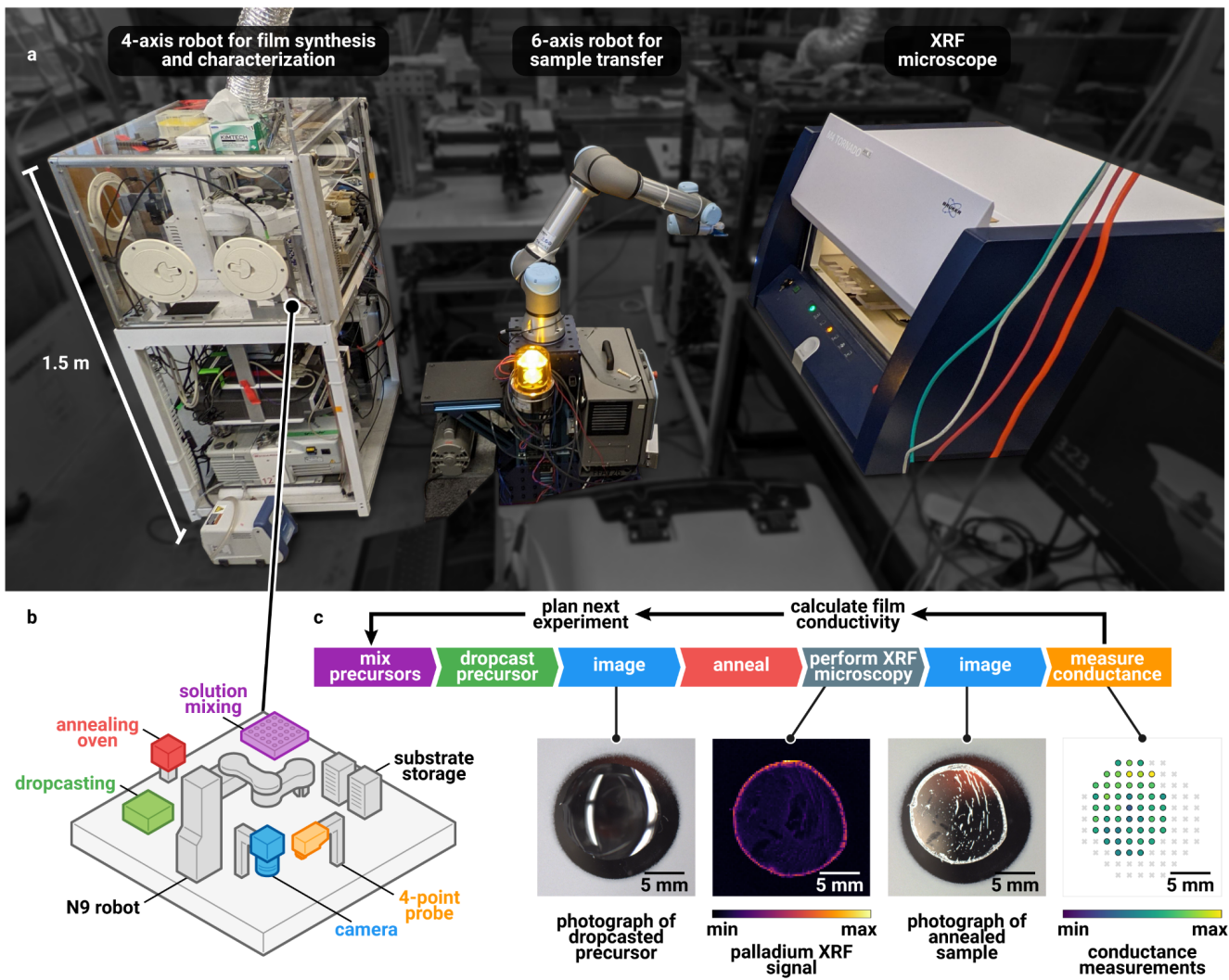

**Fig. 2 | The Ada self-driving laboratory and autonomous experimental workflow.** (**a**) Schematic of the Ada self-driving laboratory. Ada consists of two robots (N9 and UR5e) with overlapping work envelopes. These robots work together to synthesize and characterize thin film samples. The N9 robot is a 4-axis arm equipped to mix, drop cast, and anneal precursors to create thin-film samples. The N9 also performs imaging and 4-point probe conductance measurements on the films it creates. The UR5 robot is a larger 6-axis arm equipped to transport samples to additional modules, including an XRF microscope. (**b**) Steps in the automated experimental workflow. Each iteration of the experiment produces a single, drop-cast, thin-film sample, images of the sample before and after annealing, an XRF map of the quantity of palladium in the film, and a map of the film conductance measured by the 4-point-probe at different locations on the sample. After the sample is characterized, the film conductivity is calculated and the qEHVI algorithm is used to autonomously plan the next experiment.

The synthesis conditions tested during the optimization, and those conditions that created materials on the Pareto front, are highlighted in Figure 3c. The data revealed that the optimal precursors typically were those of concentrations near 6 mg mL$^{-1}$, fuel-to-oxidizer ratios below 1, and fuel blends consisting primarily of acetylacetone. Notably, our experiments did not reveal a single optimal synthesis condition. The conditions required to obtain the maximum conductivity depended in part on the annealing temperature. Specifically, conductive films created below 200 °C required precursors with predominantly acetylacetone fuel. At higher temperatures, glycine-rich fuel blends yielded samples on the Pareto front. The data shows how the fuel-to-oxidizer ratio could vary widely for fuels rich in acetylacetone yet still yield films on the Pareto front. The Pareto-optimal samples resulting from glycine-rich fuel blends, however, did not exhibit a wide range of fuel to oxidizer values. These observations highlight the richness of the data generated by the self-driving laboratory.

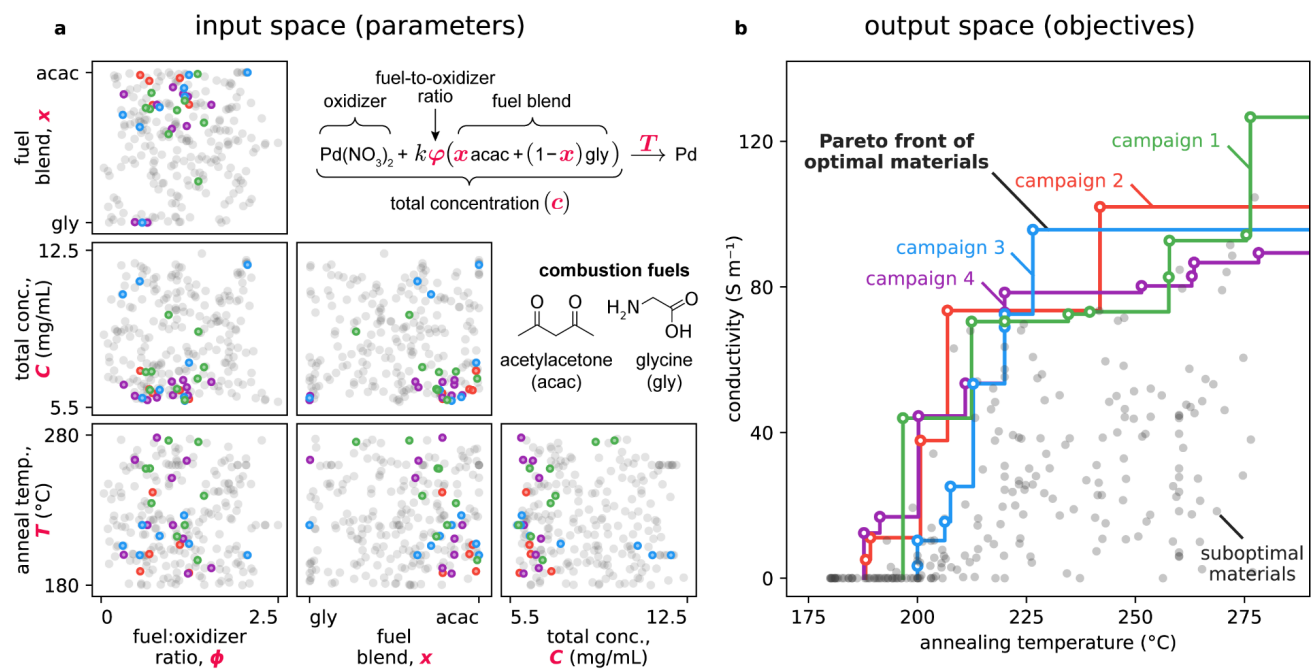

**Fig. 3 | Trade-off between annealing temperature and conductivity for the combustion synthesized palladium films.** (**a**) Maps of the combustion synthesis conditions required to obtain experimental outcomes on the Pareto front. Sampled points not on the Pareto front are shown in grey. The combustion synthesis reaction, with parameters manipulated during the optimization highlighted, is shown. In this reaction $x$ controls the fuel blend and $\varphi$ is the fuel-to-oxidizer ratio, as calculated using Jain's method[60]. The fuel-to-oxidizer ratio used is scaled by $k = 36/5(8 - 5x)$ to account for the differing reducing valences of the acetylacetone and glycine fuels (see SI). (**b**) The empirical Pareto fronts from each of the four campaigns (solid lines) reveal the observed trade-off between temperature and conductivity. The experimental points which define the fronts are shown with open markers. Sampled points not on the Pareto front are shown in grey.

We used computer simulations to quantify the benefit of the qEHVI algorithm relative to random search (an open-loop sampling technique that does not use feedback from the experiment to determine which experiment to do next). These simulations were performed by running both the random and qEHVI sampling techniques on a response surface fit to the experimental data (see methods). Scenarios with and without experimental noise were simulated by adding synthetic noise to the response surface as appropriate (see methods: "Models of the experimental response surface and noise").The hypervolume (i.e. the area under the Pareto front) was used to measure the progress of optimization (Fig. 4a). We used acceleration factor (Fig. 4b) and enhancement factor (Fig. 4c) to compare our closed-loop sampling with qEHVI to open-loop sampling with random; see methods[61]. In a noise-free scenario, qEHVI required less than 100 samples to outperform 10,000 random samples (fig. 4a). The performance of the qEHVI algorithm degraded in the presence of simulated experimental noise (see methods), but still exceeded the performance of random search. This performance decrease due to noise emphasizes the importance of minimizing experimental noise when developing a self-driving experiment. Additional benchmarks comparing other closed-loop and open-loop sampling techniques yielded similar results (see fig. S11). These findings highlight how self-driving laboratories can effectively search large materials design spaces without requiring extremely high throughput.

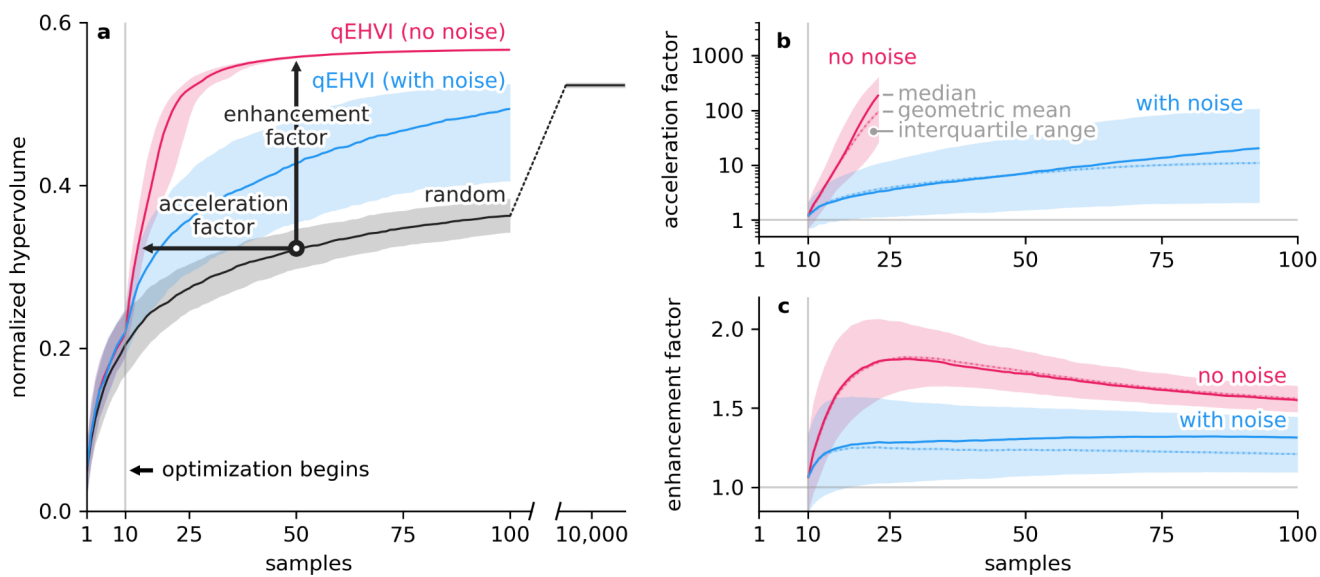

**Fig. 4 | Quantification of the benefit provided by the qEHVI algorithm in simulated optimization campaigns.** (**a**) The hypervolumes achieved by the simulated qEHVI and random searches. The median (solid line) and interquartile range (shaded bands) of the results are shown for simulations with and without simulated experimental noise. (**b**) The acceleration factors for the qEHVI algorithm relative to random search. The geometric mean is additionally shown (dashed line). (**c**) The enhancement factor for the qEHVI algorithm relative to random search.

**Translation of discovery to a scalable manufacturing process**

The practical application of combustion synthesis would require the deposition of uniform films over large areas that are inaccessible to drop casting. On this basis, we set out to combine palladium combustion synthesis with ultrasonic spray coating[62]. We sprayed precursors directly onto a preheated glass substrate[62] (Fig. 5a, methods). The precursors decomposed in less than five minutes to yield reflective, conductive palladium films (Fig. 5b). An XRF map of the films (Fig. 5c) also showed improved homogeneity relative to the drop-cast films (Fig. 2b).

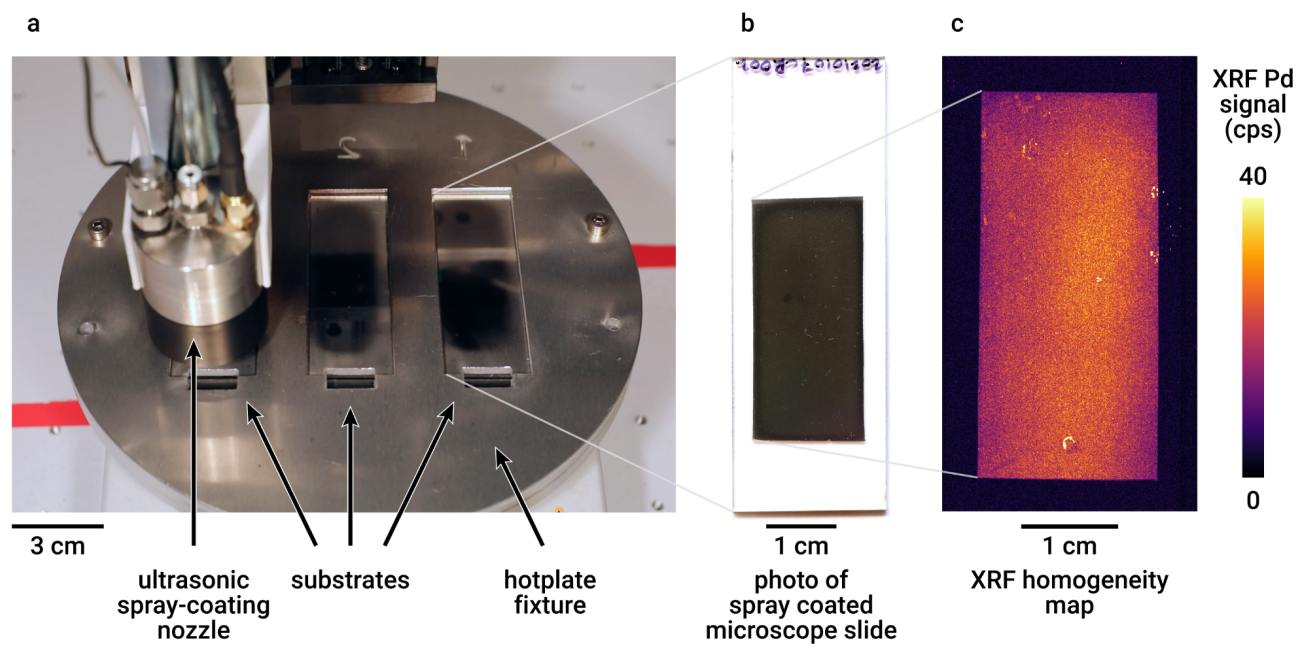

**Fig. 5 | Low-temperature spray combustion synthesis of homogeneous palladium films over large areas. (a)** Spray coating apparatus. An ultrasonic nozzle attached to an overhead XYZ stage (not visible) is used to spray palladium combustion synthesis precursors onto glass substrates placed on a hot plate with an aluminum fixture. The precursors decompose to yield palladium films. **(b)** Photograph of a typical resulting film on a 3" × 1" glass substrate. Sharp edges were produced by masking the substrate using kapton tape. **(c)** XRF map of the sample pictured in (b). To aid visualization, the photograph in panel a has been flipped horizontally.

We performed additional spray coating experiments to verify that the trends observed in the autonomous optimization translate to spray coating (i.e. that conductive palladium films can be obtained below 200 °C and that the film conductivity increases with temperature). Specifically, we spray coated palladium films using three recipes from the autonomously identified Pareto front, with temperatures of 191, 200, and 226 °C (Fig. 6, table S2). Triplicate samples were spray coated using each recipe. All three recipes yielded films approximately 50-60 nm thick, as measured by XRF microscopy (see methods). All the films were relatively uniform, with spatial variations in the film thickness less than 5% of the mean within an 8 mm × 20 mm region at the center of each sample (see methods and table S3). Spatial variations in the conductivity within the same region were measured using the robot and were less than 18% of the mean for all samples (see methods and table S3). The film conductivity of the lowest temperature recipe ($T$ = 191 °C) was $1.1 \times 10^5$ S m$^{-1}$, which is approximately 1% of the bulk conductivity of palladium. Increasing the spray coating temperature can increase the film conductivity by more than an order of magnitude. The highest temperature recipe tested ($T$ = 226 °C) yielded palladium films with a conductivity of $2.0 \times 10^6$ S m$^{-1}$ which is comparable to the conductivities of sputtered palladium films reported in the literature ($2.0 - 5.8 \times 10^6$ S m$^{-1}$; Fig. 6)[63–65]. These findings create new opportunities to deposit palladium films without vacuum onto large-area substrates, including an expanded range of temperature-sensitive polymers (e.g., Nafion[48], polyethersulfone[49], heat-stabilized polyethylene napthalate[49]). One application of this deposition process could be the fabrication of large, supported palladium membranes for more cost-effective electrocatalytic palladium membrane reactors[66].

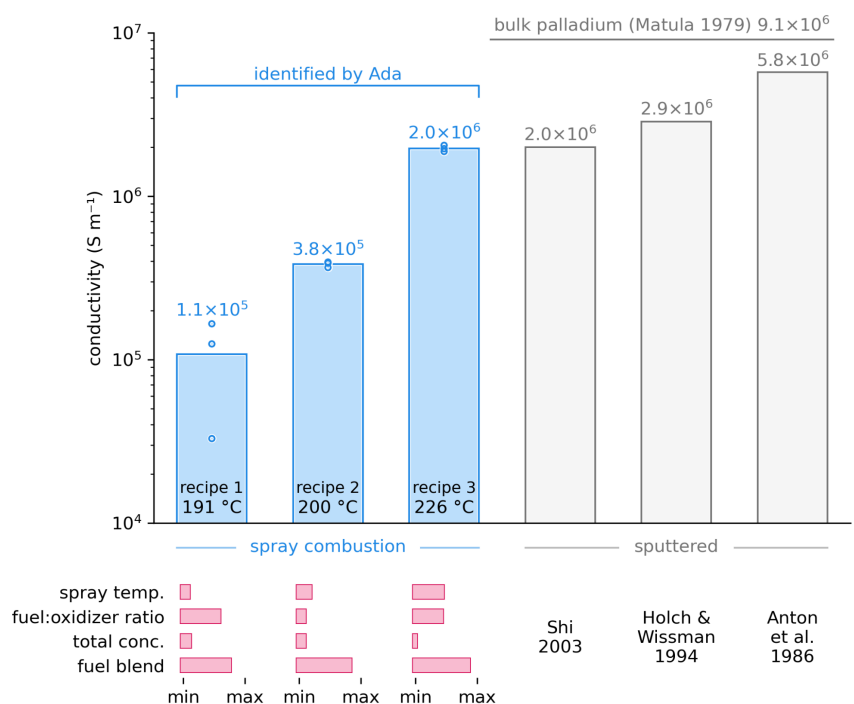

**Fig. 6 | Comparison between the conductivity of the spray-coated palladium films and sputtered films.** The conductivity values for sputtered films [63–65] and bulk palladium[67] are from previous literature. The spray coating recipes are taken directly from the Pareto front and are given in Supplementary Information Table 2.

**Conclusion**

Here we mapped out a Pareto front between film processing temperature and conductivity using a self-driving laboratory guided by the qEHVI multi-objective optimization algorithm. This tradeoff is just one example of the conflicting objectives routinely faced by materials scientists to which our method could be applied. Our approach eliminates the need for the researcher to specify preferences between competing objectives in advance of the experiment, and also produces a richer, more valuable data set. In this case, the temperature–conductivity Pareto front is more useful than optimizing conductance for a fixed temperature limit because processing temperature limits vary depending on the application. Our self-driving laboratory also identified synthesis conditions that translated to a scalable spray-coating method for depositing high-quality, high-conductivity palladium films at temperatures above 190 °C. This work shows how self-driving laboratories can potentially accelerate the translation of materials to industry, where satisfying multiple objectives is essential.

**Methods**

**Materials**

MeCN (CAS 75-05-8; high-performance liquid chromatography (HPLC) grade, ≥99.9% purity), glycine (CAS 56-40-6, ACS reagent grade, >98.5% purity) and acetylacetone (CAS 123-54-6; ≥99% purity) were purchased from Sigma-Aldrich. Urea (CAS 57-13-6, ultra-pure; heavy metal content 0.01 ppm) was purchased from Schwarz/Mann. Palladium(II) nitrate hydrate ($Pd(NO_3)_2 \cdot H_2O$; Pd ~40% m/m; 99.9% Pd purity, CAS 10102-05-3) was purchased from Strem Chemicals, Inc. All chemicals were used as received without further purification.

**Manual preparation of stock solutions**

The self-driving laboratory is provided with starting materials in the form of stock solutions which are prepared manually and then placed in capped 2 mL HPLC vials in a tray where they can be accessed by the self-driving laboratory. All solutions were prepared at a concentration of 12 mg mL$^{-1}$. The Pd(NO$_3$)$_2$•H$_2$O solution was prepared using MeCN as a solvent while all other solutions were prepared using deionized H$_2$O.

**Preparation of glass substrates and other consumables**

In addition to stock solutions, the self-driving laboratory uses consumable glass substrates (75 mm × 25 mm × 1 mm microscope slides; VWR catalogue no. 16004-430), 2 mL HPLC vials (Canadian Life Science), and 200 μL pipettes (Biotix, M-0200-BC). These are placed in appropriate racks and trays for access by the robotics.

The HPLC vials and pipettes were used as received, whereas the microscope slides were cleaned by sequential sonication in detergent, deionized water, acetone, and isopropanol as previously described[7]. 18 mm diameter wells were then created on the microscope slides using a sprayed enamel coating (DEM-KOTE enamel finish) and circular masks placed at the center of each slide (Supplementary Fig. 1). The wells serve to confine the precursor solution before it dries.

**Self-driving laboratory**

The self-driving laboratory consists of a precision 4-axis laboratory robot (N9, North Robotics) coupled with a 6-axis collaborative robot (UR5e, Universal Robotics). The 4-axis robot is equipped to perform entire thin film deposition and characterization workflows and is described in our previous work[7]. The 6-axis robot enables samples to be transferred to a variety of additional modules, including the XRF microscope used here. Both robots are equipped with vacuum-based tools for substrate handling. All robots and instruments were controlled by a PC with software written in Python.

**Overview of autonomous robotic workflow**

The majority of operations in the autonomous robotic workflow are performed by the 4-axis laboratory robot. Samples are transported between the 4-axis robot and the XRF microscope by the 6-axis robot.

The 4-axis robot prepared each sample by combining stock solutions to form a precursor mixture, drop casting this precursor onto a glass slide, and then annealing the sample in a forced convection oven (Supplementary Fig. 1). The samples were then characterized by white light photography before and after annealing, X-ray fluorescence microscopy, and 4-point-probe conductance measurements. The resulting data was then automatically analyzed using a custom data pipeline implemented in Python. Finally, the result of the experiment was fed to a Bayesian optimizer which used an expected hypervolume improvement acquisition function to select the next experiment to be performed. Each of these steps is described in further detail below.

**Autonomous workflow step 1: mix precursors**

The 4-axis robot formulated each precursor by pipetting varying volumes of the stock solutions described above into a clean 2-mL HPLC vial. Gravimetric feedback from an analytical balance (ZSA120, Scientech) was used to minimize and record pipetting errors. The precursor was mixed by repeated aspiration and dispensing.

**Autonomous workflow step 2: drop cast precursor**

The 4-axis robot used a vacuum-based substrate handling tool to place a clean glass slide onto a tray. This robot then created a thin film sample by using a pipette to drop cast 98 μL of the precursor into a predefined well on the slide. The solution was ejected from the pipette at a rate of 5 μL s$^{-1}$ from a height of approximately 1.5 mm above the top surface of the substrate.

**Autonomous workflow step 3: image precursor droplet**

The 4-axis robot acquired visible-light photographs of each sample before annealing. This robot positioned samples 90 mm below a camera (FLIR Blackfly S USB3; BFS-U3-120S4C-CS) using a Sony 12.00 MP CMOS sensor (IMX226) and an Edmund Optics 25 mm C Series Fixed Focal Length Imaging Lens (#59-871). The C-mount lens was connected to the CS-mount camera using a Thorlabs CS- to C-Mount Extension Adapter, 1.00"-32 Threaded, 5 mm Length (CML05). The sample was illuminated from the direction of the camera using a MIC-209 3-W ring light. For imaging, the lens was opened to f/1.4, and black flocking paper (Thorlabs BFP1) was placed 10 cm behind the sample.

**Autonomous workflow step 4: annealing**

After drop casting, the 4-axis robot used the substrate handling tool to transport the precursor-coated slide into a purpose-built miniature convection oven for annealing at a variable temperature between 180 °C and 280 °C. The most important features of the oven are a low-thermal mass construction (lightweight aluminum frame with glass-fibre insulation) and internal and external fans. These features enable rapid heating and cooling of the sample. A pneumatically actuated lid enables robotic access to the sample. The oven employs a ceramic heating element (P/N 3559K23, McMaster Carr) controlled by a PID temperature controller (P/N CN7523, Omega Engineering). A type-K thermocouple located in the oven air space provides temperature feedback to the controller. In the experiments performed here, the sample was inserted into the oven which was then ramped at 40 °C per minute to the temperature set point, which was then held for 450 s. Upon completion of the hold, the oven lid was opened and a cooling fan turned on to blow ambient temperature air through the oven and over the sample. The sample was removed from the oven after the temperature dropped below 60 °C. The oven was further cooled to below 40 °C prior to loading of the next sample.

**Autonomous workflow step 5: X-Ray fluorescence hyperspectral imaging and data analysis**

The self-driving laboratory acquired hyperspectral X-ray fluorescence (XRF) images of each sample using a Bruker M4 TORNADO X-ray fluorescence microscope equipped with a customized sample fixture. Samples were transported to the XRF microscope by the UR5e 6-axis robotic arm equipped with a vacuum-based substrate handling tool similar to the one used by the 4-axis N9 robot. A dedicated exchange tray accessible to both robots enabled samples to be passed from one robot to the other.

The XRF microscope has a rhodium X-ray source operated at 50 kV / 600 µA / 30 W and polycapillary X-ray optics yielding a 25 µm spot size on the sample. The instrument employs twin 30 mm$^2$ silicon SSD detectors and achieves an energy resolution of 10 eV. Hyperspectral images were taken over a 20 mm × 20 mm area at a resolution of 125 × 125 pixels. The XRF spectra obtained (reported in counts) were scaled by the integration time (50 ms) and the energy resolution (10 eV) to yield units of counts s$^{-1}$ eV$^{-1}$.

To quantify the relative amount of palladium in the film, the palladium Lyman-alpha X-ray fluorescence line (2.837 keV) was integrated from 2.6 to 3.2 keV. The resulting counts were converted to film thickness estimates by applying a calibration factor obtained using reference samples (see below). 97 points of interest are defined within the XRF hypermap of the sample, as defined in Supplementary Figure 3. For each point of interest, the average XRF counts per second were calculated over a 3 mm × 3 mm area.

**Autonomous workflow step 6: image annealed film**

The self-driving laboratory acquired visible light photographs (as described in step 3) of each sample after annealing.

**Autonomous workflow step 7: measure film conductance and robustly estimate film conductivity**

After hyperspectral XRF imaging, the sample was returned by the UR5e robot to the N9 robot for film conductance measurements. Four-point probe conductance measurements were performed with a Keithley Series K2636B System Source Meter instrument connected to a Signatone four-point probe head (part number SP4-40045TBN; 0.040-inch tip spacing, 45 g pressure, and tungsten carbide tips with 0.010-inch radii) by a Signatone triax to BNC feedthrough panel (part number TXBA-M160-M). The source current was stepped from 0 to 1 mA in 0.2 mA steps. After each current step, the source meter was stabilized for 0.1 s and the voltage across the inner probes was then averaged for 3 cycles of the 60 Hz power line (i.e. for 0.05 s) and recorded. Conductance measurements were made on the same 97 points of interest as analyzed in the XRF data, as defined in Supplementary Figure 3.

The film conductivity was calculated using a custom data analysis pipeline implemented in Python using the open-source Luigi framework[68]. This pipeline combined conductance data and XRF data to estimate the film conductivity at each of the 97 points of interest on the sample.

For each set of current-voltage measurements at each position on each sample, the RANSAC robust linear fitting algorithm[69] was used to extract the conductance (dI/dV). The voltage compliance limit of the K2636B was set to 10 V and voltage measurements greater than 10 V were therefore considered to have saturated the Source Meter instrument and automatically discarded by the data analysis pipeline.

The conductivity of the thin films was then calculated by combining the 4-point-probe conductance data with the film thicknesses estimated by XRF:

$$\sigma = \frac{\ln 2}{\pi} \times \frac{dI}{dV} \times t^{-1} \quad (1)$$

where dI/dV is the conductance from the 4-point-probe measurement, $t$ is estimated film thickness from the XRF measurements, and $\sigma$ is conductivity.

Due to the poor morphology of the drop-cast films, a robust conductivity estimation scheme was employed. First, conductance data was excluded for any measurement positions with zero conductance. Next, outliers were excluded from the remaining conductance data using a kernel density exclusion method (see below). Outliers were also excluded from the XRF film thickness estimates using the same filtering method. Conductivities were calculated for each position on the sample for which neither conductance nor XRF data was excluded. The mean of these conductivities was returned to the optimizer (see below). In cases where all points were discarded, a mean conductivity of 0 was reported.

The outlier kernel density exclusion method was performed by calculating Gaussian kernel density estimates for the conductance and XRF data, normalizing the density between 0 and 1, and rejecting data points with a kernel density below 0.3. Bandwidths of $5\times10^{-3}$ $\mu\Omega^{-1}$ $m^{-1}$ and $5\times10^{3}$ cps were used for the conductance and XRF data, respectively.

**Autonomous workflow step 8: experiment planning using the qEHVI algorithm**

The experiment parameters for each optimization experiment performed on the autonomous laboratory were determined by the qEHVI[55] multiobjective Bayesian optimization algorithm. In brief, this algorithm proposes experiments expected to increase the area underneath the Pareto front by the largest amount. More formally, the algorithm proposes a batch of $q$ experiments ($q = 1$ here, but $q$ could be increased to exploit parallelized experimentation) which are collectively expected to increase the hypervolume between the Pareto front and a reference point by the largest amount. The hypervolume is a generalization of the concept of volume for an arbitrary number of dimensions; this generalization supports optimization with more than two objectives. The reference point must be specified prior to the optimization and specifies a minimum value of interest for each objective.

The algorithm involves two major conceptual steps: modelling the objectives from data and proposing the next experiment. In the configuration used here, each objective is assumed to be independent and is modelled with an independent gaussian process (see Supplementary Figs. 5-6). Based on the models for each objective, the expectation value of the hypervolume improvement associated with any candidate experiment can be computed; the candidate experiment with the largest expected hypervolume improvement is selected. We ran the qEHVI algorithm using the implementation available in the open-source BoTorch Bayesian optimization library[70,71]. We used a temperature reference point at the upper limit of the experiment (280 °C) so that any outcome with a processing temperature below this upper limit would be targeted. We used a dynamic conductivity reference point set to 5% of the running observed maximum conductivity. This dynamic reference point ensured that the optimization would identify Pareto-optimal outcomes over a wide range of conductivity values and did not require prior knowledge of the scale of conductivity values expected. We used heteroskedastic Gaussian processes to model both the conductivity and the temperature[70]. We assigned each conductivity point an uncertainty equal to 20% of its value, which is comparable to the repeatability of the experiment. Zero uncertainty was assigned to the temperature values, which were manipulated rather than responding variables and were trivial to model.

**Calibration of XRF signal against reference samples of known thickness**

To enable palladium film thickness to be estimated from the XRF signal, a calibration procedure was performed on sputtered palladium reference samples having four different nominal thicknesses (10, 50, 100, 250 nm). These samples were characterized by profilometry and XRF. A linear relationship between the film thickness and the XRF counts was observed (see Supplementary Fig. 2). This relationship was used to estimate the thickness of each sample from the XRF data.

The reference samples were sputtered onto clean glass microscope slides (see cleaning procedure above) using a Univex 250 sputter deposition system with a DC magnetron source at 100 W and an argon working pressure of $5\times10^{-6}$ bar. The deposition chamber base pressure is $5\times10^{-9}$ bar. Films were deposited after 1 min of presputtering. The substrate holder rotated at 10 rpm. Nominal film thickness was monitored using a quartz crystal microbalance mounted in the sputter chamber. A 2-inch diameter palladium sputter target was used (99.99%, ACI Alloys). Step edges for profilometry were obtained by placing strips of Kapton™ tape onto the substrates prior to sputtering and removing these after sputtering. The substrates were rinsed with acetone and IPA to remove any Kapton™ tape residue prior to performing profilometry.

Profilometry was performed on the reference samples using a Bruker DektakXT stylus profilometer. XRF was performed on the reference samples robotically using the same settings used for the drop-casted samples during the optimization campaigns (see above).

**Deposition and characterization of spray-coated samples**

The spray coater was built from an ultrasonic nozzle (Microspray, USA) mounted to a custom motorized XYZ gantry system (Zaber Technologies Inc., Canada) above a hot plate (PC-420D, Corning, USA). Precursor ink was fed to the nozzle by a syringe pump (cavro centris pump PN: 30098790-B, Tecan Trading AG, Switzerland). The ultrasonic spray nozzle was operated at 3 W and 120 kHz. For each recipe, a total of 700 µL of precursor was sprayed onto a glass substrate (75 mm × 25 mm × 1 mm microscope slides; VWR catalogue no. 16004-430) placed on a custom aluminum fixture mounted to the hotplate. The approximate substrate temperatures were measured using a thermocouple attached to a glass substrate with thermal cement. This instrumented substrate was placed at a position on the hotplate fixture symmetrical to the position where the substrates to be coated were placed. The hotplate power was adjusted until the steady-state temperature of the instrumented substrate was within 4 °C of the desired temperature before spray coating each of the recipes reported here. To achieve consistent thermal contact between the substrates and the hotplate fixture, both the instrumented substrate and substrate to be coated were affixed to the hotplate with thermal paste (TG-7, Thermaltake Technology Co., Taiwan). The spray coater nozzle speed was 5.1 mm s$^{-1}$, the nozzle-to-substrate distance was 15 mm, the spray flow rate was 2 µL s$^{-1}$, and the carrier gas flow rate was 7 L min$^{-1}$. When spraying, the nozzle moved in a serpentine pattern consisting of twelve 50-mm lines with 25 mm spacing (see an illustration of the pattern in fig. S14). The coating on each sample was produced by repeating this spray pattern three times with no delay between passes. After spray coating, the samples were left to anneal on the hot plate for five minutes.

The spray-coated palladium films were characterized at 26 locations on a 2 × 13 grid within an 8 × 20 mm region of interest at the center of the film (see fig. S14). The amount of palladium at each location was measured using the XRF microscope and converted to a film thickness estimate by applying the same calibration method used for the drop-cast films. The film conductance at each location was measured using the 4-point probe system on the robot described above. The film conductivity was calculated at each of the 26 measurement locations by combining the 4-point probe conductance and XRF film thickness values for that location using equation (1). The mean and standard deviations of the 26 resulting thickness and conductivity values are reported for each film in table S3.

**Computer simulations of optimization algorithm performance**

Computer simulations were used to study the performance of the qEHVI algorithm for optimizing the combustion synthesis experiments. A model of the experimental response surface was built from the experimental data. Experimental optimizations were then simulated by sampling the model using grid search, random search, Sobol sampling, and the qEHVI and qParEGO algorithms.. Optimization performance was quantified with and without simulated experimental noise. The performance of qEHVI relative to random sampling was quantified using the acceleration factor (AF) and enhancement factor (EF) metrics[61]. These simulation procedures are described in more detail below.

**Models of the experimental response surface and noise**

Gaussian process regression was used to create a model of the experimental response surface using the combined data from all four optimization campaigns. This model predicts the experimental

23/35

outputs (i.e. annealing temperature and conductivity) from the experimental inputs (i.e. fuel-to-oxidizer ratio, fuel blend, total concentration, and annealing temperature). Our model is composed of two separate Gaussian processes, as implemented by the scikit-learn Python package[69] . Each Gaussian process is regressed on a single experimental output (i.e. either temperature or conductivity) and all four of the experimental inputs. The kernels used for the conductivity ($k_\text{cond}$) and temperature ($k_\text{temp}$) models are:

$$k_\text{cond}(x, x') = k_\text{lin}(x, x') \times k_\text{SE}(x, x') + k_\text{noise}(x, x') \quad (2)$$

$$k_\text{temp}(x, x') = k_\text{lin}(x, x')^* \times k_\text{SE}(x, x')^* \quad (3)$$

$$k_\text{noise}(x, x') = noise \text{ if } x = x' \text{ else } 0 \quad (4)$$

where $k_\text{lin}$ is a constant kernel, $k_\text{SE}$ is a squared exponential kernel, and $k_\text{noise}$ is a white noise kernel, and $*$ indicates that the lengthscale of the kernel is fixed to 1.

The four types of input data and two types of output data were each normalized prior to training the model. To simplify the optimization to be strictly a maximization problem, the temperature values (which must be minimized) were multiplied by negative one. The leave-one-out cross-validation residuals (LOOCV; Supplementary Fig. 4; Supplementary Table 4) are comparable to the measured experimental uncertainties (Supplementary Table 1). We also plotted the LOOCV residuals as a function of each input (Fig. S6), each modeled output (Fig. S7), and sampling order (Fig. S8) and observed that the distribution of the residuals was largely random.

For the simulations with no noise, the model posterior means were used directly to represent the experiment. For simulations with experimental noise, noisy conductivity values were simulated by randomly sampling a modified Maxwell-Boltzmann distribution. First, the Maxwell-Boltzmann distribution was flipped across the *y*-axis by negating the *x* term. Second, the mean of this Maxwell-Boltzmann distribution was set to the noiseless model posterior mean. Finally, the variance of the Maxwell-Boltzmann distribution was set to be equal to the noise level (or variance) of the white noise kernel. We chose to employ Maxwell-Boltzmann noise to model the experimental noise because of the tendency of drop-casted samples to exhibit a wide range of downwards deviations in the apparent conductivity due to the poor sample morphology.

The Maxwell-Boltzmann probability density function is

$$P_B(x) = \sqrt{\frac{\pi}{2}} \frac{x^2 \exp\left(\frac{-x^2}{2a^2}\right)}{a^3} \quad (5)$$

where $a$ is the distribution parameter. Since we set the variance of Maxwell-Boltzmann distribution ($\sigma_B^2$) equal to the white noise kernel noise level ($\sigma_{\text{noise}}^2$),

$$\sigma_{\text{noise}}^2 = \sigma_B^2 = \frac{a^2(3\pi - 8)}{\pi} \quad (6)$$

We computed the distribution parameter for the Maxwell-Boltzmann-distributed simulated experimental noise as:

$$a = \sqrt{\frac{\pi \sigma_{\text{noise}}}{3\pi - 8}} \quad (7)$$

If subtracting the Maxwell-Boltzmann noise from the posterior mean resulted in a value less than zero, the noisy model value was set to zero.

**Sampling strategies**

To compare the performance of closed-loop and open-loop approaches, several sampling strategies (grid search, random search, Sobol sampling, and the qEHVI and qParEGO algorithms) were used to sample both the noise-free and noisy experimental models. Random sampling was performed by generating samples from a uniform distribution across the entire normalized input space. Sobol sampling was performed with a scrambling technique such that each Sobol sequence is unique to yield a statistically meaningful distribution of optimizations[72,73]. qEHVI and qParEGO are initiated with 10 scrambled Sobol points. The qEHVI algorithm was configured as for the physical experiments detailed above. The qParEGO algorithm was configured using the default settings, except for the reference point which was configured in the same way done for qEHVI. The complete benchmarking results are shown in Figure S11.

**Simulated optimization campaigns**

The performance of each sampling strategy (grid, random, Sobol, qParEGO, and qEHVI) was determined both with and without experimental noise. Each simulated optimization campaign was performed for 1000 replicates and 100 experimental iterations, except for random which was performed for 100,000 iterations for use in the acceleration calculations (shown in fig. 4b). If an optimization algorithm produced an error during optimization, then that replicate was removed and repeated.

**Pareto front and hypervolume**

The Pareto front is defined by the set of samples for which no other sample simultaneously improves all the objectives. When assessing the performance of a simulation, the hypervolume was computed using the least desirable value of each objective function (the nadir objective vector) as the reference point (i.e. zero conductivity and 280 °C temperature). This reference point was held constant for evaluating all of the simulated optimization campaigns. We assessed the performance of the noisy optimizations such that the only difference between the noisy and noise-free optimizations was the information provided to the optimization algorithm. To perform this assessment, we followed the procedure described by Bakshy and coworkers[59] wherein the hypervolume for optimizations using the noisy experimental model were calculated from the equivalent point from the noiseless model. Since each objective of the model is normalized to the range [0, 1], the hypervolume of the model is in the range [0, 1]. For each simulated optimization campaign, the normalized hypervolume was calculated at each iteration. When calculating acceleration and enhancement factors, each of the 1000 simulated qEHVI campaigns was compared to each of the 1000 simulated random campaigns, resulting in 1,000,000 comparisons.

**Calculation of acceleration and enhancement factor**

The acceleration factor quantifies how much faster one sampling technique is than another (equation 5). For example, if sampling technique B requires 40 samples to reach the performance attained by technique A after 20 samples, the acceleration factor of A relative to B at 20 samples is 2.

$$AF_{A:B}(n_a) = \frac{n_b}{n_a}$$

$$\text{s.t. } P_B(n_b) \geqslant P_A(n_a), \ \min n_b \quad \textbf{(8)}$$

where $AF_{A:B}(n_a)$ is the acceleration of technique A with respect to B at $n_a$ samples, and $P_x(n)$ is the performance of technique X at $n$ samples. Note that it is possible for sampling technique A to outperform B such that there exists no value of $n_b$ where $P_B \geqslant P_A$. In these cases, more samples with technique B are required to make the comparison, otherwise $AF_{A:B}$ is not calculable. If $AF_{A:B}$ is not calculable, then a lower bound acceleration factor is calculated by assuming that the slow sampling technique would beat the fast sampling technique if it observed one more sample. The acceleration factor in fig. 4b was reported until these lower bound estimates compose more than 25% of all of the acceleration comparisons.

The enhancement factor of one sampling technique with respect to another for a given number of samples is defined as the ratio of their performance values for the same number of observations (equation 6). For example, if sampling technique A reaches a performance value of 7 after 20 samples, and technique B reaches a performance value of 2 after 20 samples, the enhancement of technique A is 3.5 at 20 samples.

$$EF_{A:B}(n) = \frac{P_A(n)}{P_B(n)} \quad (9)$$

where $EF_{A:B}(n)$ is the acceleration factor of sampling technique $A$ with respect to $B$ after $n$ samples. When $P_A(n) = 0$ and $P_B(n) = 0$, then $EF_{A:B}(n) = 1$. When $P_A(n) > 0$ and $P_B(n) = 0$, then $EF_{A:B}(n)$ is not calculable. To compare the AF and EF from the repeated simulations, the median, geometric mean and interquartile range were calculated.

**Acknowledgements**

The authors are grateful to Natural Resources Canada's Energy Innovation Program (EIP2-MAT-001) for financial support. The authors are grateful to the Canadian Natural Science and Engineering Research Council (RGPIN-2018-06748), Canadian Foundation for Innovation (229288), Canadian

Institute for Advanced Research (BSE-BERL-162173), and Canada Research Chairs for financial support. B. P. M., F. G. L. P., T. D. M., and C. P. B. acknowledge support from the SBQMI's Quantum Electronic Science and Technology Initiative, the Canada First Research Excellence Fund, and the Quantum Materials and Future Technologies Program. We would like to acknowledge the many open-source software communities without whose efforts this project would not have been possible. Please see the supplementary information for further details.

**Data availability**

The raw data recorded by the self-driving laboratory during the experiments reported here are available at https://github.com/berlinguette/ada. Additional data related to this paper may be requested from the authors.

**Author contributions**

C. P. B. conceived and supervised the project. B. P. M., K. E. D., and F. G. L. P. designed and performed the autonomous optimization experiments. M. B. R., K. O., C. W., K. E. D., O. P., M. S. E., B. P. M., and F. G. L. P. developed the robotic hardware. M. S. E., O. P., and K. E. D. developed and configured the robotic control software. F. G. L. P. and T. H. H. developed the data analysis software with input from N. T., K. E. D., and B. P. M., F. G. L. P., M. M., and B. P. M. performed and analyzed the simulations. M. S. E. and B. P. M. configured the EVHI optimization algorithm and interfaced it with the self-driving laboratory. K. O., H. N. C., and C. C. R. developed the spray coater hardware and software. C. C. R. performed the spray coating experiments. N. T. performed additional data analysis. D. J. D. performed additional experiments. All authors participated in the writing of the manuscript.

# Advancing the Pareto front for thin-film materials using a self-driving laboratory

## *Supplementary information*

Benjamin P. MacLeod[1,2,*], Fraser G. L. Parlane[1,2,*], Connor C. Rupnow[1,2,3], Kevan E. Dettelbach[1], Michael S. Elliott[1], Thomas D. Morrissey[1,2], Ted H. Haley[1], Oleksii Proskurin[1], Michael B. Rooney[1], Nina Taherimakhsousi[1], David J. Dvorak[2], Hsi N. Chiu[1], Christopher E. B. Waizenegger[1], Karry Ocean[1], Mehrdad Mokhtari[1] & Curtis P. Berlinguette[1,2,3,4,†]

[1]Department of Chemistry, The University of British Columbia, 2036 Main Mall, Vancouver, BC V6T 1Z1, Canada.
[2]Stewart Blusson Quantum Matter Institute, The University of British Columbia, 2355 East Mall, Vancouver, BC V6T 1Z4, Canada.
[3]Department of Chemical and Biological Engineering, The University of British Columbia, 2360 East Mall, Vancouver, BC V6T 1Z3, Canada.
[4]Canadian Institute for Advanced Research (CIFAR), MaRS Centre, 661 University Avenue Suite 505, Toronto, ON M5G 1M1, Canada.

*These authors contributed equally to this work.
†Corresponding author. Email: cberling@chem.ubc.ca

**Outline**

# Supplementary tables

## Supplementary Table 1 | Initialization points used for the optimizations

| sample | fuel blend ($x$) | anneal temperature ($T$, °C) | fuel to oxidizer ratio ($\phi$) | concentration ($C$, mg mL$^{-1}$) | conductivities across campaigns 1-4 mean (S m$^{-1}$) | std. dev. (S m$^{-1}$) |
|---|---|---|---|---|---|---|
| 1 | 0.5 | 220 | 2.0 | 0.006 | 21 | 12 |
| 2 | 0.7 | 260 | 0.5 | 0.012 | 31 | 16 |
| 3 | 1.0 | 200 | 2.0 | 0.012 | 8 | 2 |
| 4 | 0 | 220 | 0.5 | 0.006 | 68 | 8 |
| 5 | 0.1 | 260 | 1.0 | 0.012 | 37 | 11 |
| 6 | 0.7 | 200 | 0.5 | 0.012 | 1 | 1 |
| 7 | 0.9 | 220 | 1.0 | 0.006 | 67 | 7 |
| 8 | 1.0 | 260 | 2.0 | 0.012 | 17 | 10 |

## Supplementary Table 2 | Recipes from the Pareto front chosen for spray coating

| recipe | fuel blend ($x$) | anneal temperature ($T$, °C) | fuel-to-oxidizer ratio ($\phi$) | concentration ($C$, mg mL$^{-1}$) |
|---|---|---|---|---|
| 1 | 0.78 | 191 | 1.56 | 0.0066 |
| 2 | 0.85 | 200 | 0.32 | 0.0065 |
| 3 | 0.90 | 226 | 1.17 | 0.0060 |

## Supplementary Table 3 | Spray-coated sample characterization results

| spray-coated sample recipe | thickness, mean (nm) | thickness, std. dev. (nm) | thickness, std dev. / mean (%) | conductivity, mean (S m$^{-1}$) | conductivity, std. dev. (S m$^{-1}$) | conductivity, std. dev. / mean (%) |
|---|---|---|---|---|---|---|
| 1 | 49.2 | 2.2 | 4.5 | $3.3 \times 10^4$ | $6 \times 10^3$ | 17.4 |
|   | 50.5 | 2.4 | 4.8 | $1.25 \times 10^5$ | $1.3 \times 10^4$ | 11.0 |
|   | 50.6 | 2.0 | 4.0 | $1.7 \times 10^5$ | $2 \times 10^4$ | 12.1 |
| 2 | 61.1 | 1.4 | 2.3 | $3.89 \times 10^5$ | $1.1 \times 10^4$ | 2.8 |
|   | 59.0 | 1.6 | 2.7 | $4.0 \times 10^5$ | $3 \times 10^4$ | 8.0 |
|   | 59.7 | 2.1 | 3.5 | $3.97 \times 10^5$ | $8 \times 10^3$ | 2.0 |
| 3 | 48.3 | 1.4 | 2.9 | $1.88 \times 10^6$ | $8 \times 10^4$ | 4.4 |
|   | 54.2 | 0.9 | 1.7 | $1.95 \times 10^6$ | $3 \times 10^4$ | 1.6 |
|   | 51.9 | 1.6 | 3.1 | $2.05 \times 10^6$ | $6 \times 10^4$ | 2.8 |

## Supplementary Table 4 | Leave-One-Out Cross-Validation analysis of simulation models

| model | RMSE (S m$^{-1}$) | NRMSE (range; %) | MAE (S m$^{-1}$) | $r^2$ | bias (S m$^{-1}$) |
|---|---|---|---|---|---|
| **Single campaigns** | | | | | |
| campaign 1 (2020-12-18_17-38-40) | 10.6 | 10.4 | 7.699 | 0.836 | 0.092 |
| campaign 2 (2020-12-23_17-06-50) | 13.8 | 15.5 | 10.961 | 0.693 | -0.016 |
| campaign 3 (2021-01-04_08-37-39) | 10.7 | 11.2 | 7.671 | 0.828 | 0.287 |
| campaign 4 (2021-01-12_16-26-56) | 11.6 | 9.2 | 7.238 | 0.871 | 0.336 |
| **Combined campaigns** | | | | | |
| all campaigns (no noise) | 12.4 | 9.8 | 9.116 | 0.800 | -0.123 |
| all campaigns (with noise) | 16.3 | 12.8 | 12.064 | 0.659 | 1.820 |

# Supplementary materials and methods

## Calculation of the fuel to oxidizer ratio for the combustion synthesis reaction

An idealized form for the combustion synthesis reaction studied here is:

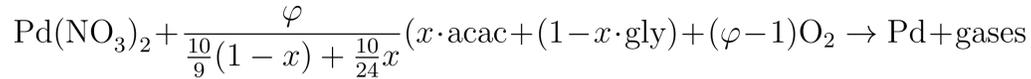

$$\mathrm{Pd(NO_3)_2} + \frac{\varphi}{\frac{10}{9}(1-x) + \frac{10}{24}x}(x \cdot \mathrm{acac} + (1-x \cdot \mathrm{gly})) + (\varphi - 1)\mathrm{O_2} \rightarrow \mathrm{Pd} + \mathrm{gases}$$

This assumes that the available oxidizers and fuels react fully with each other, and with additional atmospheric oxygen as needed to yield pure, metallic palladium. This idealized reaction is based on Jain's method (Jain, Adiga, and Pai Verneker 1981) which calculates an overall fuel-to-oxidizer ratio on the basis of the oxidizing or reducing valence of each species involved in a combustion reaction. The reducing valences used are +4 for carbon, +1 for hydrogen, 0 for nitrogen, -2 for oxygen and $+v$ for a metal forming a compound in which the metal has formal charge $v$ (e.g. +2 for Pd).

## Transformations from normalized optimizer variables to dimensional variables used by the robotics

As shown in Fig. 2, the optimizations performed here were cast in terms of normalized variables:

- The fuel to oxidizer ratio, $\varphi$ (dimensionless)
- The fuel blend, $x$ (dimensionless)
- The total precursor concentration, $C$ (mg/mL)
- The annealing temperature, $T$ (°C)

The values for these variables were chosen by the qEHVI optimization algorithm before each experiment. While the annealing temperature $T$ chosen by the algorithm could be passed directly to the robotic hardware, the other variables ($\varphi$, $x$, and $C$) required transformation into quantities suitable for execution by the robot. Specifically, a transformation was applied to these quantities to determine the required volumes of the $\mathrm{Pd(NO_3)_2}$, glycine, and acetylacetone stock solutions and water diluent required to form the precursor ink for each experiment. The transformations required that the total volume of ink required ink be specified (200 µL for the present experiments) and are expressed in terms of the following quantities:

- The normalized variables introduced above ($\varphi$, $x$, and $C$)
- The total volume of ink to be mixed ($V_\mathrm{ink} = 200 \mu L$)
- The required volume of $\mathrm{Pd(NO_3)_2}$ solution ($V_\mathrm{Pd}$)
- The required volume of glycine solution solution ($V_\mathrm{gly}$)
- The required volume of acetylacetone solution ($V_\mathrm{acac}$)
- The required volume of water used as diluent ($V_\mathrm{diluent}$)
- The reducing valences and molar masses of each compound involved in the combustion, as well as the concentrations of the associated stock solutions:

| Compound | reducing valence | molar mass (g mol⁻¹) | stock solution concentration (mg mL⁻¹) |
|---|---|---|---|
| Pd(NO$_3$)$_2$ | $R_{Pd}$ = -10 | $MM_{Pd}$ = 230.43 | $C_{Pd} = C_{stocks}$ = 12 |
| glycine | $R_{gly}$ = 9 | $MM_{Gly}$ = 75.07 | $C_{Gly} = C_{stocks}$ = 12 |
| acetylacetone | $R_{acac}$ = 24 | $MM_{Acac}$ = 100.13 | $C_{Acac} = C_{stocks}$ = 12 |

The volumes of chemicals used are found by solving the following equations numerically:

$$\frac{V_{gly}}{V_{Pd}} = \phi(1-x)\left(\frac{R_{Pd}}{R_{gly}}\right)\left(\frac{M_{gly}}{M_{Pd}}\right)$$

$$\frac{V_{acac}}{V_{Pd}} = \phi(x)\left(\frac{R_{Pd}}{R_{acac}}\right)\left(\frac{M_{acac}}{M_{Pd}}\right)$$

$$V_{ink} = V_{Pd} + V_{gly} + V_{acac} + V_{diluent}$$

$$\frac{V_{ink} - V_{diluent}}{V_{ink}} = \frac{C}{C_{stocks}}$$

**Manual screening experiments**

We qualitatively compared the decomposition temperatures of palladium combustion synthesis precursors using manual screening experiments. In these experiments, precursors containing glycine, urea, or acetylacetone as the fuel were drop cast onto glass substrates and allowed to dry in air. The dried precursors were then placed on a hotplate preheated to a specified temperature and observed by eye. Metallic films were never obtained for hotplate temperatures below 180 °C. The precursors containing glycine or acetylacetone exhibited a change in appearance earlier than those containing urea and were found to yield conductive films after annealing on a hotplate set to 350 °C. Conductive films were obtained from the urea-containing precursors only upon further heating.

## Supplementary figures

All figures containing numerical data were created in Python using the matplotlib library.

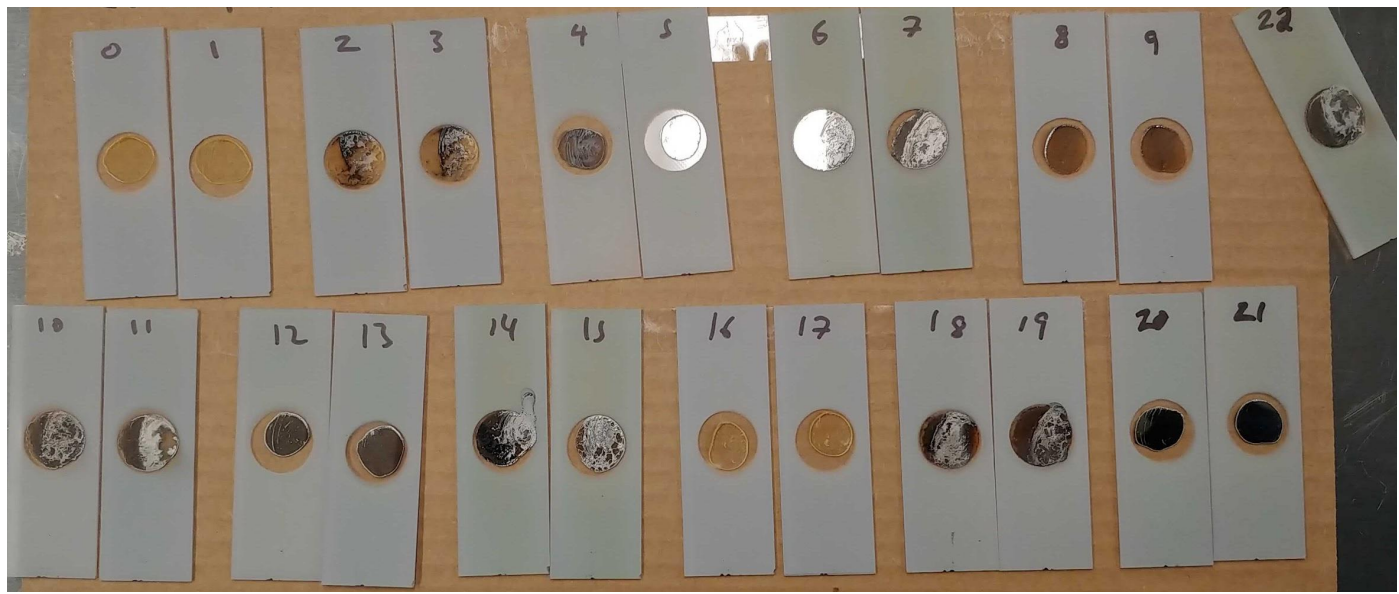

**Figure S1** | Photographs of typical drop casted films created by the robotic laboratory. The substrates are glass microscope slides. A grey spray paint which is poorly wet by the precursors was used to define an 18-mm diameter circular well at the center of each slide. The precursors are dispensed into this well.

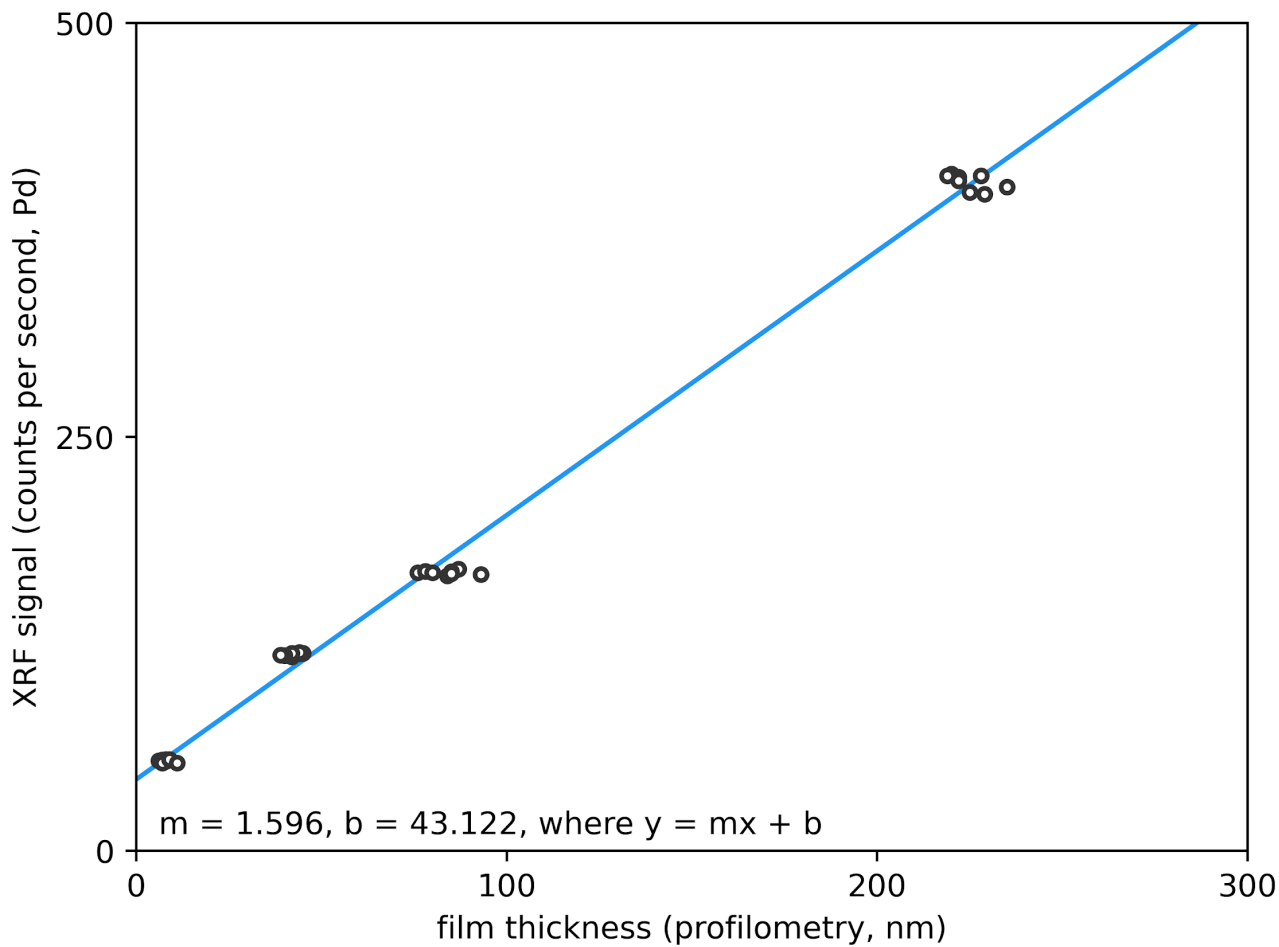

**Figure S2** | Characterization of sputtered palladium films used for XRF calibration. Films of nominal thickness 10 nm, 50 nm, 100 nm, and 250 nm were deposited by sputtering and then characterized using XRF and profilometry. A linear relationship was observed between film thickness and XRF intensity.

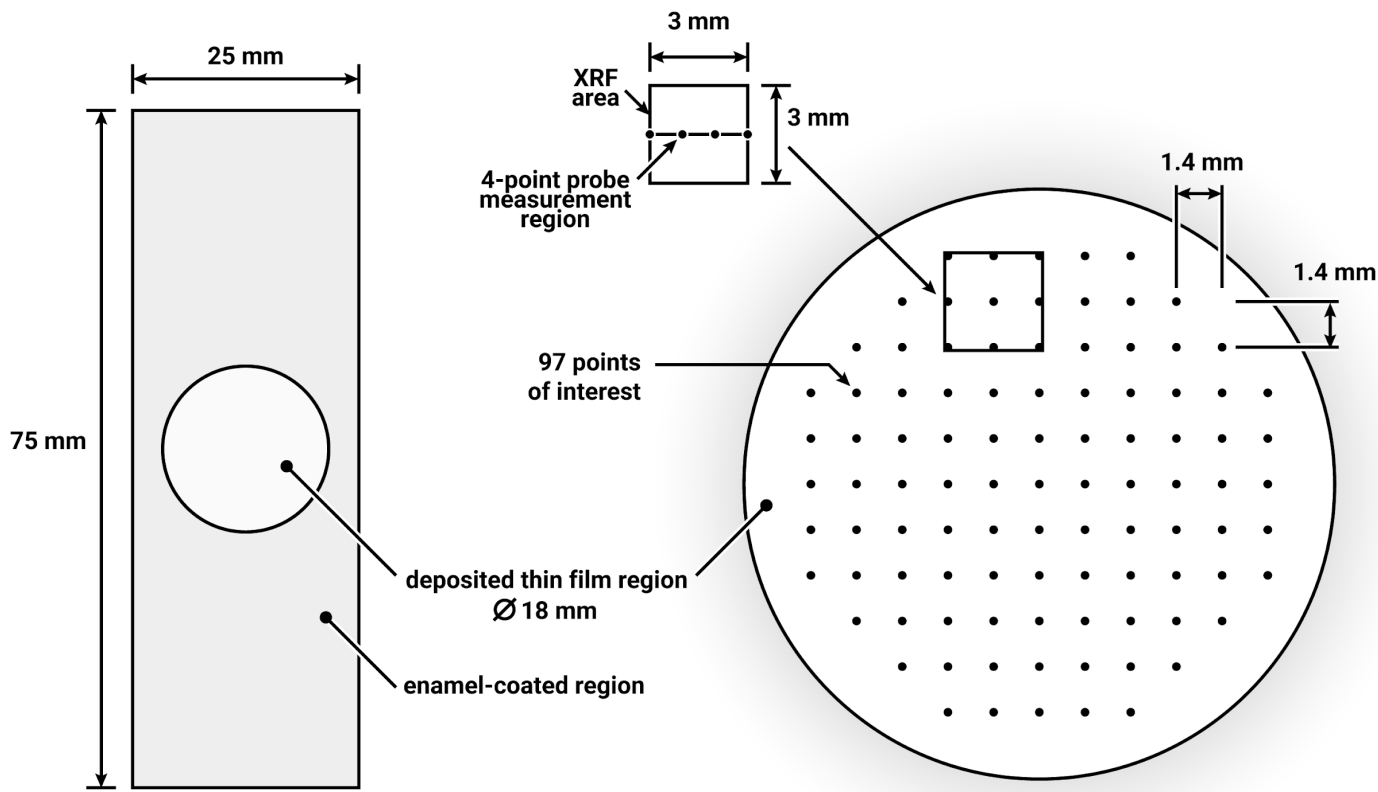

**Figure S3** | Layout of the microscope slide sample and measurement regions. Each microscope slide has an enamel coating that masks out a region on the slide with a diameter of 18 mm. The palladium thin film is deposited in this region. The conductance is measured with a 4-point probe at each point of interest. The XRF signal for palladium is measured over a 3 mm by 3 mm region centered on each point of interest.

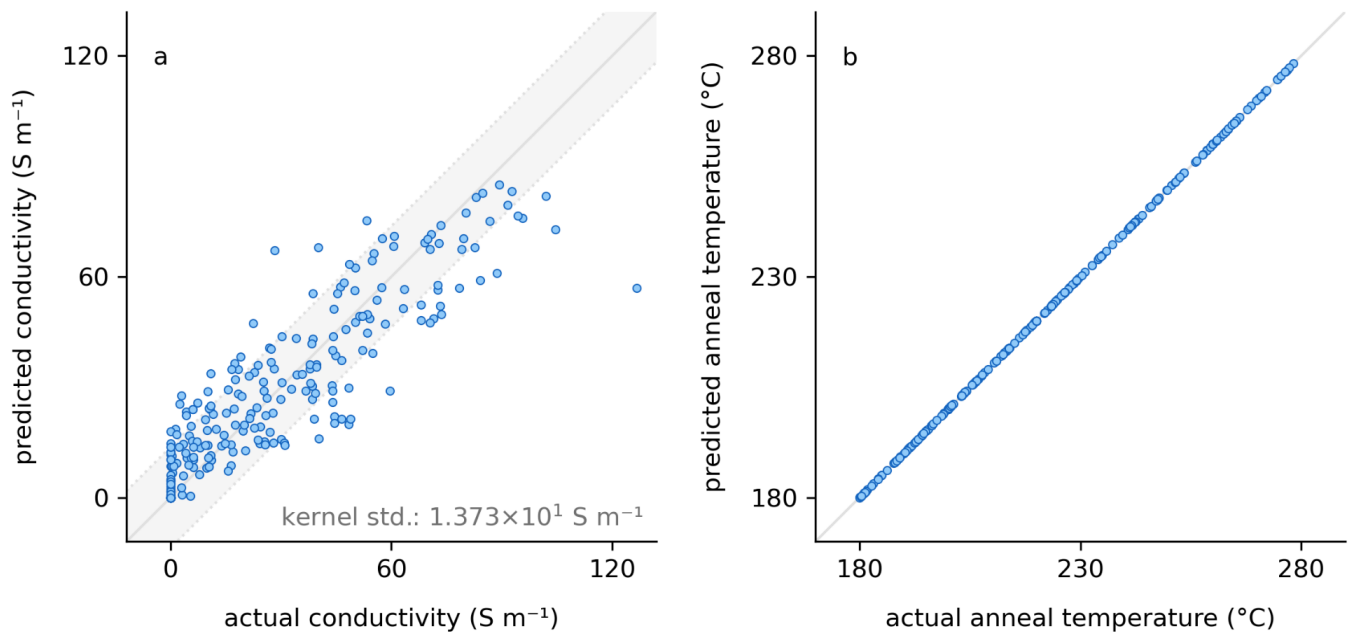

**Figure S4 | Leave-One-Out Cross-Validation (LOOCV) analysis of the noise-free model of the experimental response surfaces for (a) conductivity and (b) temperature.** The model was created by training a Gaussian process model on all the combined data from all four optimization campaigns. (See methods.) The amplitude of the experimental noise in the conductivity training data was estimated using a white noise kernel in the Gaussian process regression model. This noise estimate is plotted as a grey band spanning ±1 standard deviation.

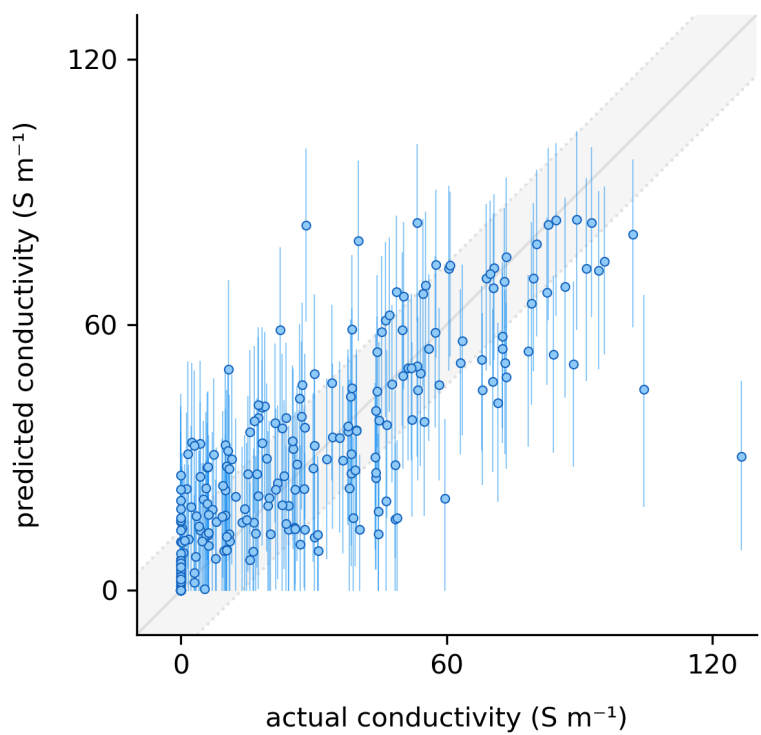

**Figure S5 | Leave-One-Out Cross-Validation (LOOCV) analysis of the noisy model of the experimental response surface.** The median (blue dots) and interquartile range (blue bars) of the noise model is reported by sampling the noise model $1\times10^6$ times at each point in the LOOCV analysis. The amplitude of the experimental noise in the conductivity training data was estimated using a white noise kernel in the Gaussian process regression model. This noise estimate is plotted as a grey band spanning ±1 standard deviation.

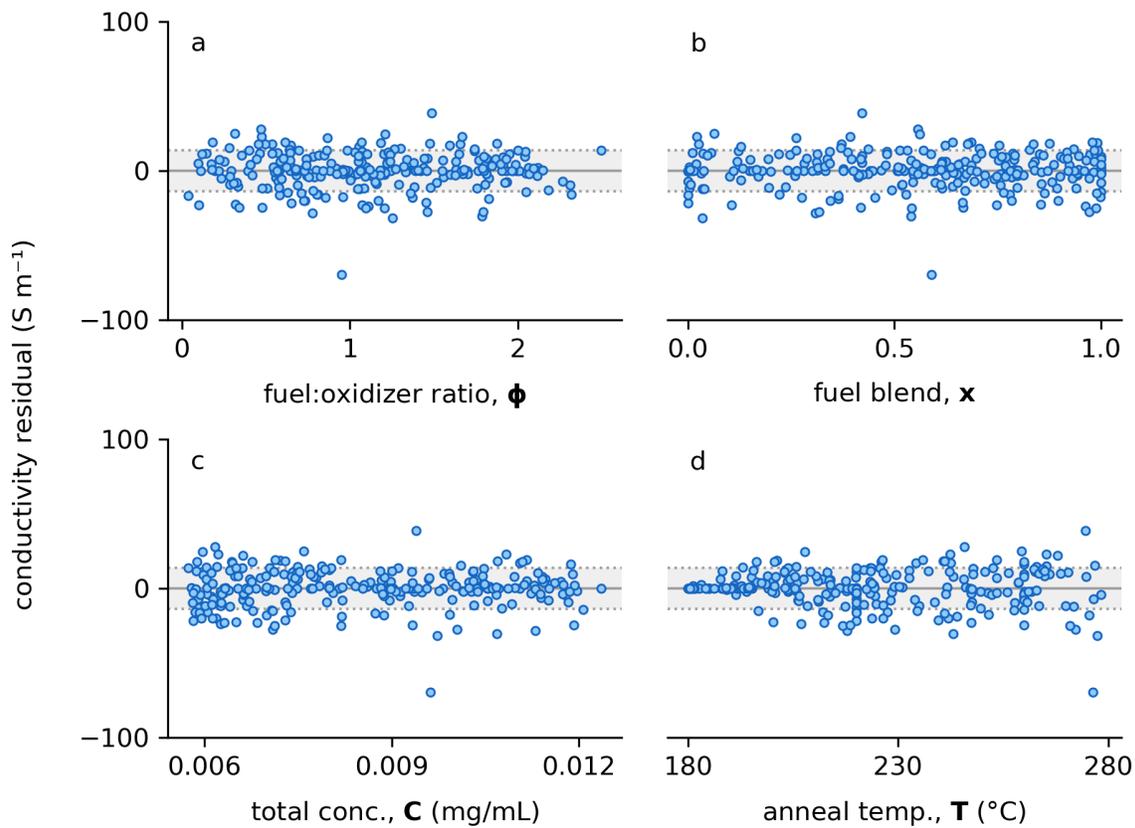

**Figure S6 | Model residuals as a function of model inputs.** The conductivity residuals from the Leave-One-Out Cross-Validation (LOOCV) analysis are plotted as a function of the input parameters revealing that the model uncertainty has little structure across the input space. The amplitude of the experimental noise in the conductivity training data was estimated using a white noise kernel in the Gaussian process regression model. This noise estimate is plotted as a grey band spanning ±1 standard deviation.

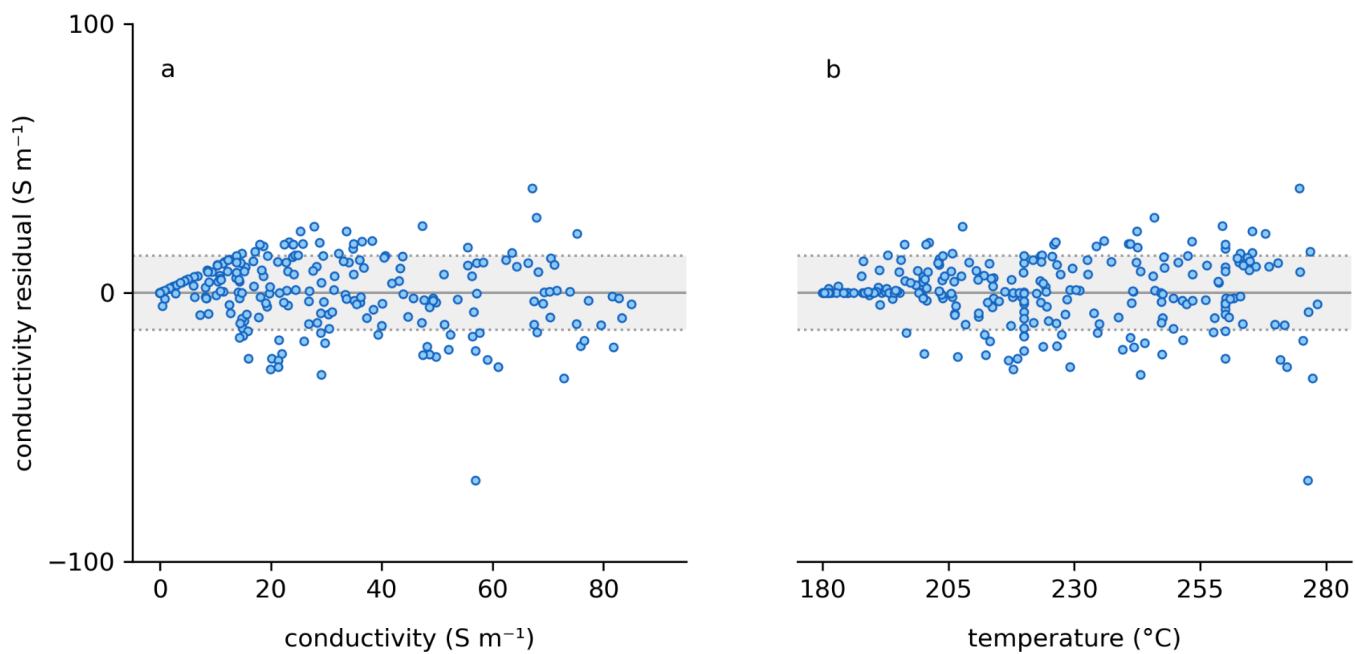

**Figure S7 | Model residuals as a function of model outputs.** The conductivity residuals from the Leave-One-Out Cross-Validation (LOOCV) analysis are plotted as a function of the model outputs. Note that the structure seen in panel a from 0 to 20 S m$^{-1}$ is a result of predicted conductivities that are negative being clipped to 0 S m$^{-1}$. The amplitude of the experimental noise in the conductivity training data was estimated using a white noise kernel in the Gaussian process regression model. This noise estimate is plotted as a grey band spanning ±1 standard deviation.

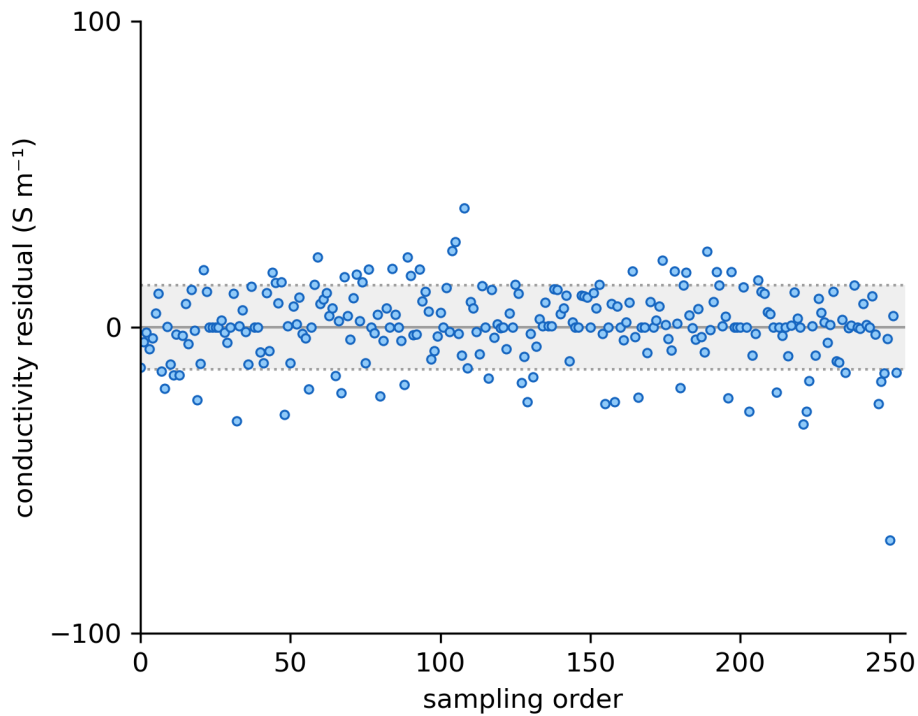

**Figure S8 | Model residuals as a function of sampling order.** The conductivity residuals from the Leave-One-Out Cross-Validation (LOOCV) analysis are plotted as a function of the sampling order. The amplitude of the experimental noise in the conductivity training data was estimated using a white noise kernel in the Gaussian process regression model. This noise estimate is plotted as a grey band spanning ±1 standard deviation.

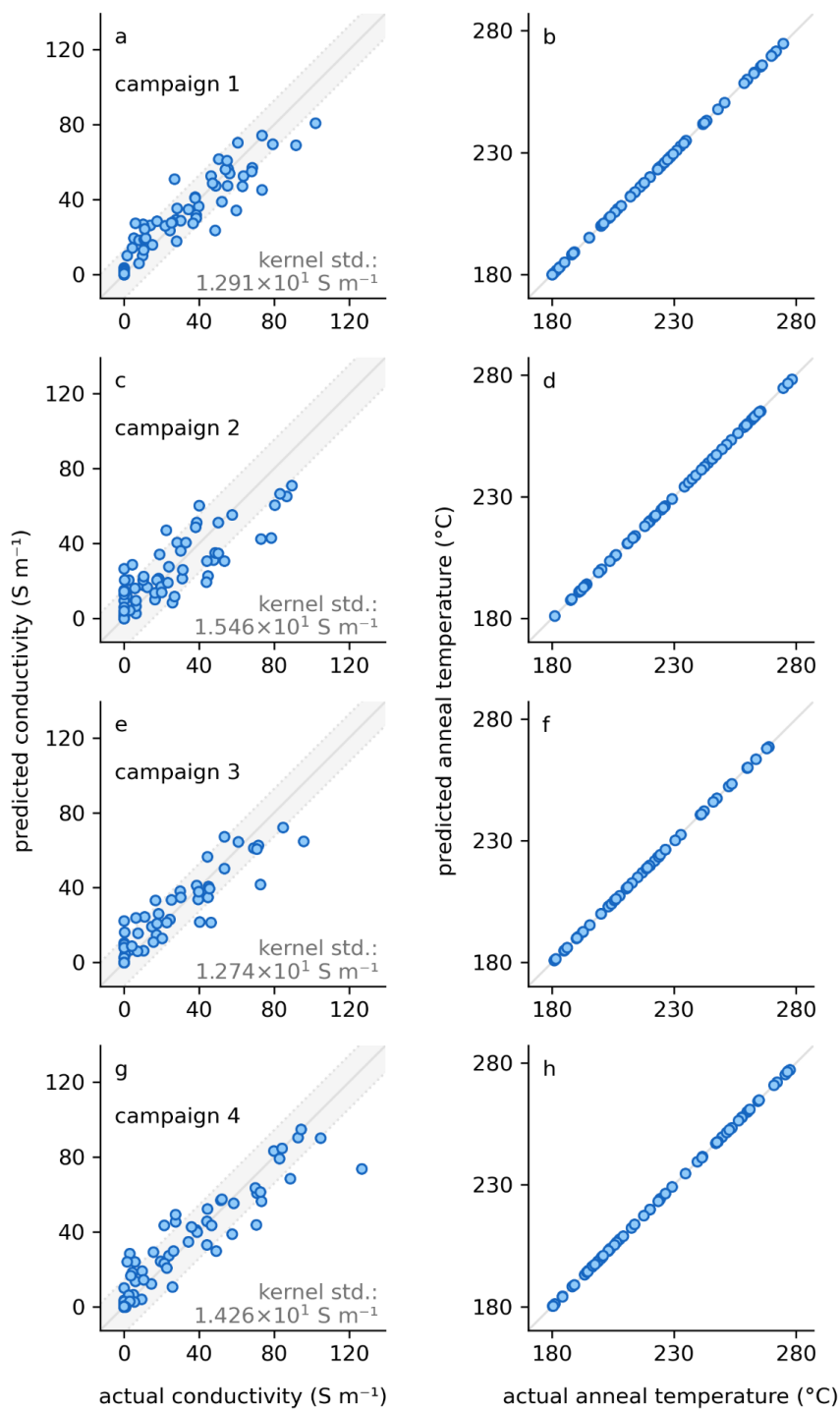

**Figure S9 | Leave-One-Out Cross-Validation (LOOCV) analysis of each of the four experimental campaigns.** The amplitude of the experimental noise in the conductivity training data was estimated using a white noise kernel in the Gaussian process regression model. The RMSE and noise estimate of the conductivity model was calculated for each experimental campaign. This noise estimate is plotted as a grey band spanning ±1 standard deviation (a, c, e, and g).

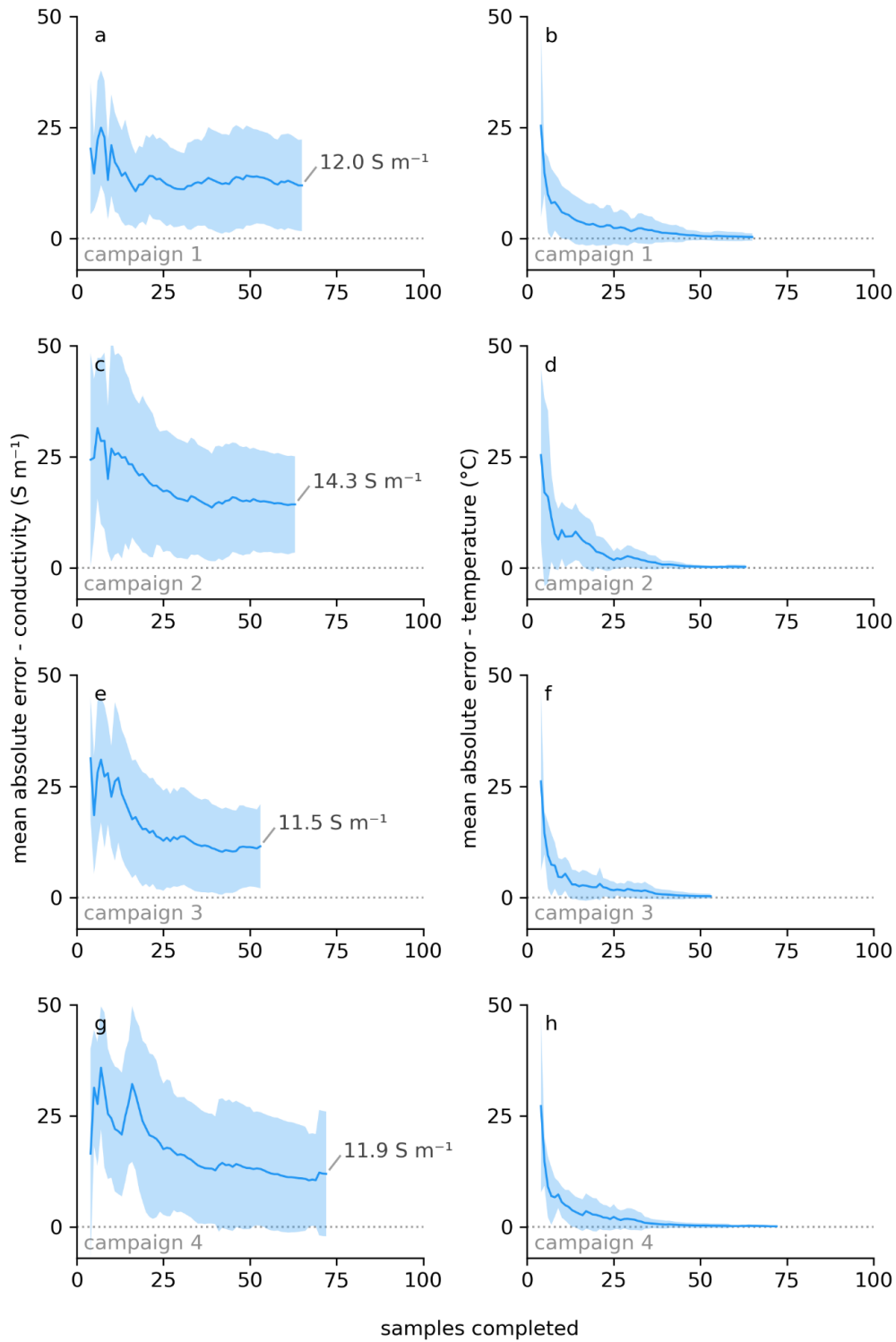

**Figure S10 | Leave-One-Out Cross-Validation (LOOCV) analysis of each of the four experimental campaigns at each sample.** The LOOCV was computed after each sample was observed. The mean and standard deviation of the mean absolute error is plotted as a function of the number of samples observed.

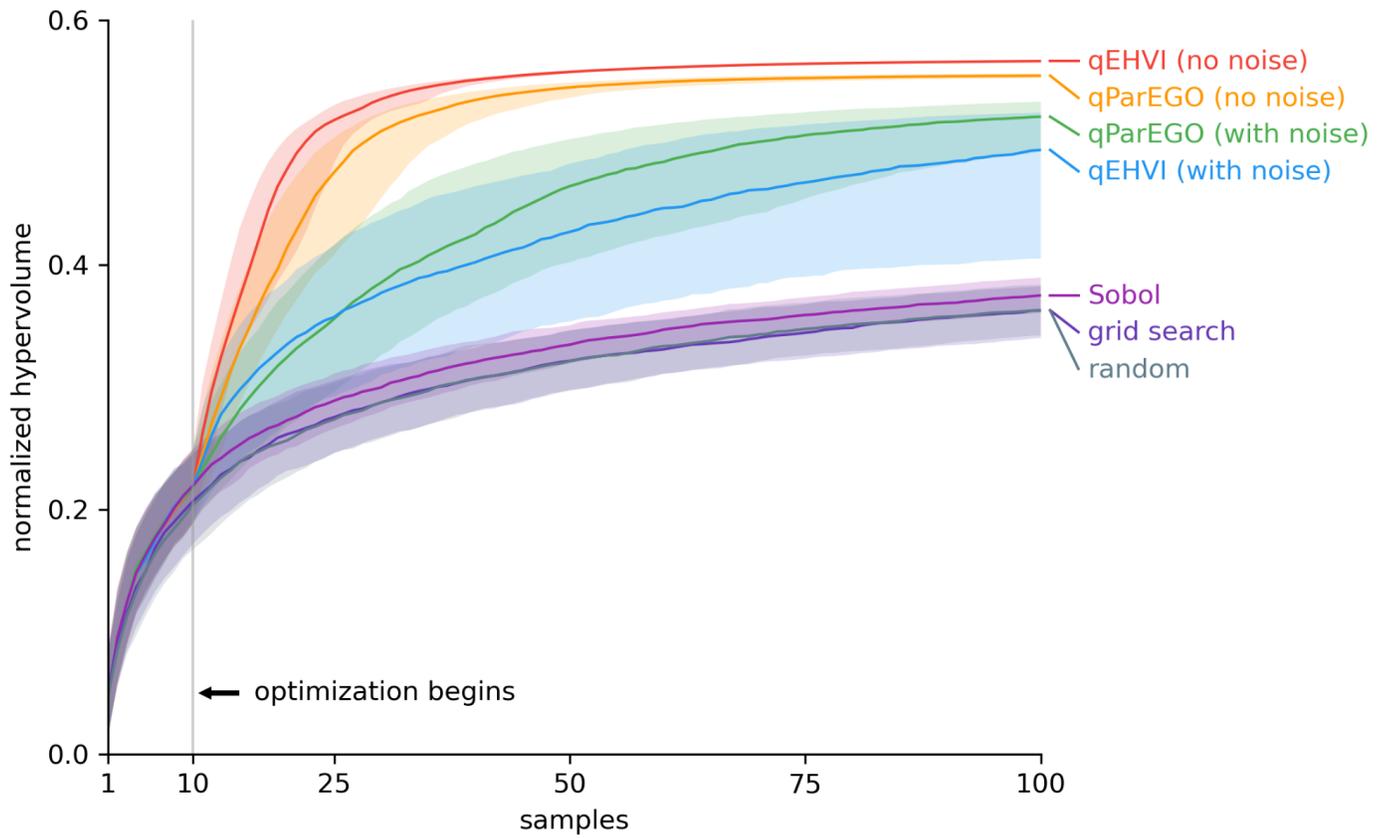

**Figure S11 | Benchmarking of qEHVI against alternative multi-objective sampling strategies.** The performance of the qEHVI algorithm is compared to alternative multi-objective sampling strategies (namely the qParEGO algorithm, grid search, random sampling, and Sobol sampling) in simulated optimization campaigns, both with and without experimental noise.

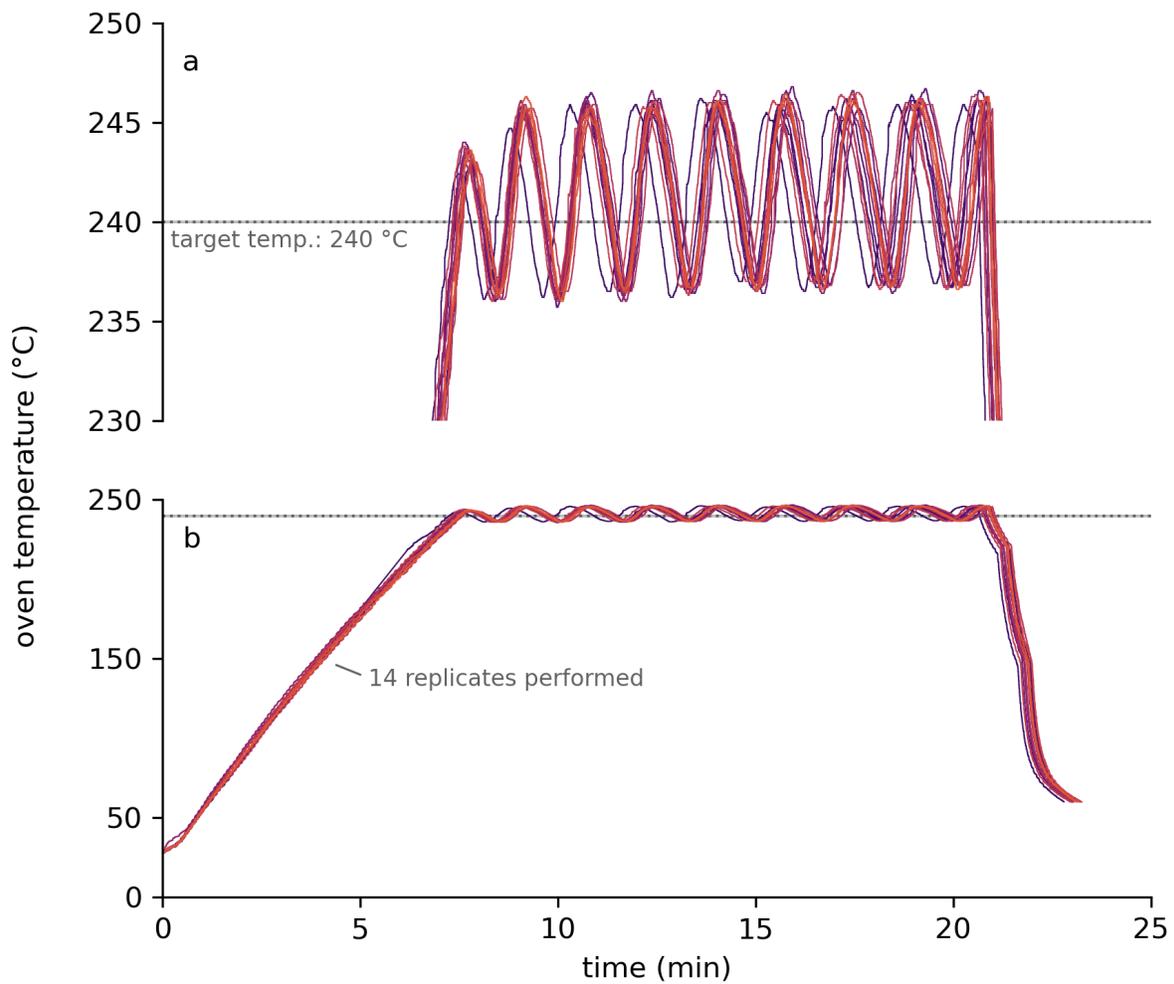

**Figure S12 | Oven temperature repeatability analysis.** The temperature of the annealing oven was recorded as a function of time when the oven was set to 240 °C. This experiment was repeated 14 times.

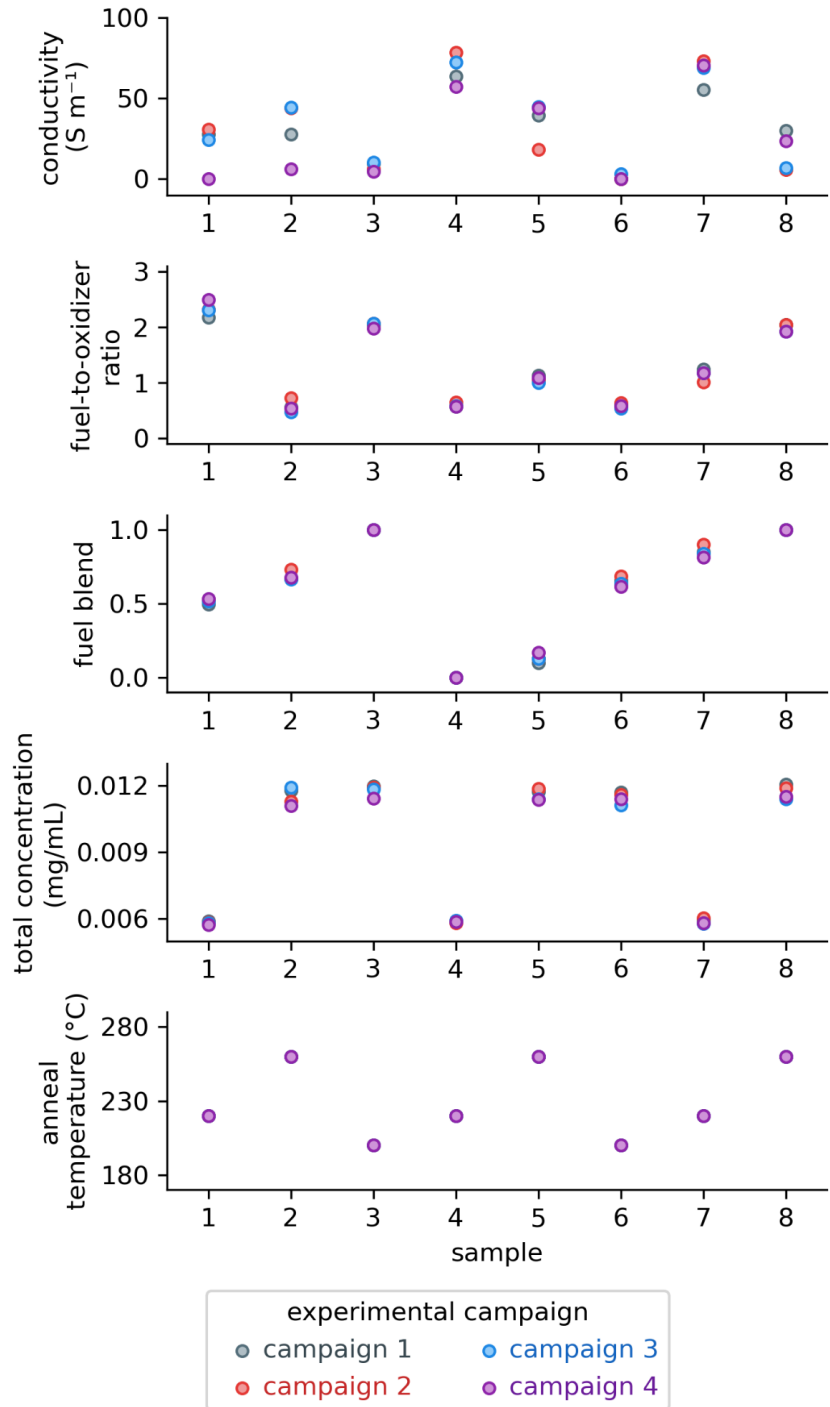

**Figure S13 | Repeatability of experimental recipe conditions and conductivity measurements.** The first eight experiments of each of the four campaigns have the same targeted experimental conditions.

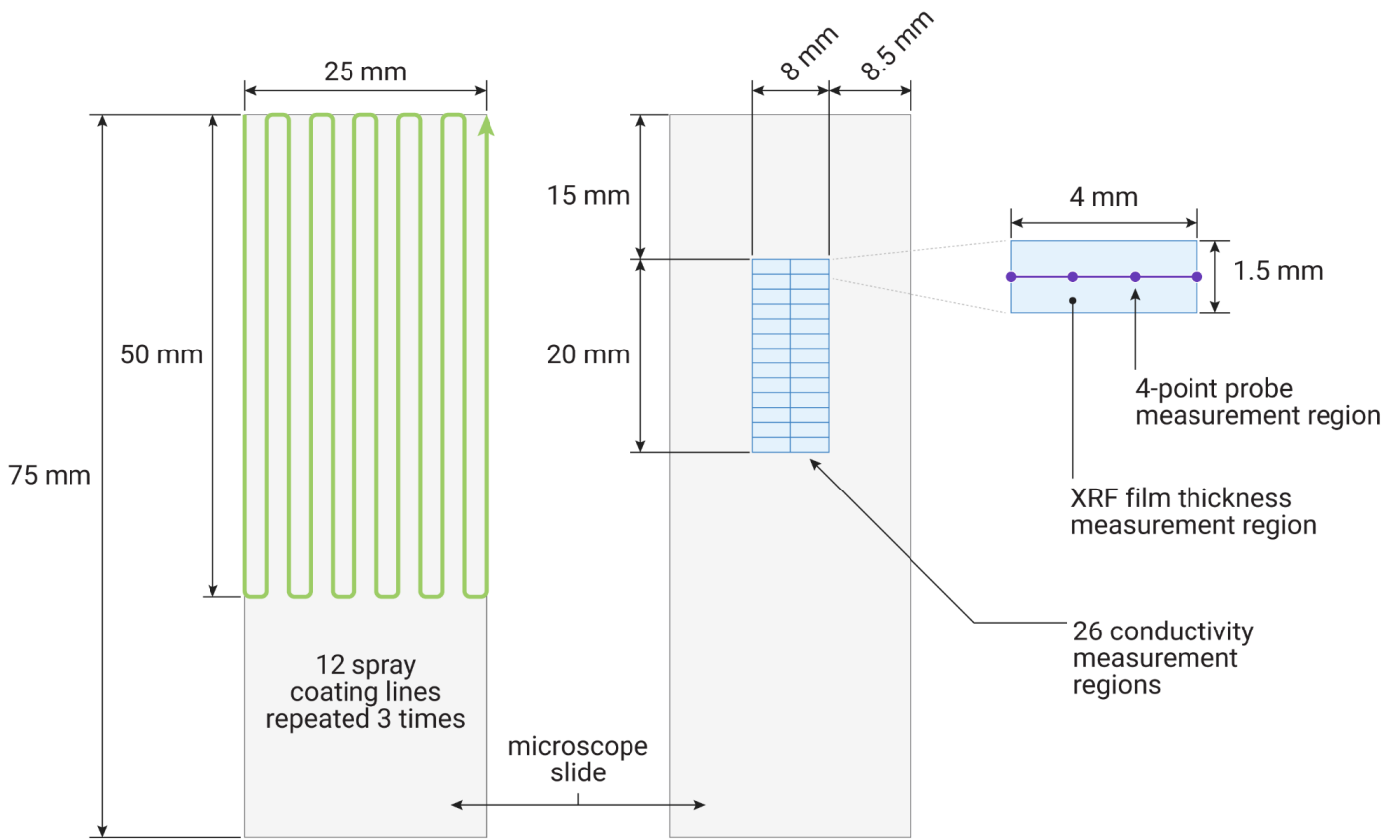

**Figure S14 | Spray coating pattern and conductivity measurement region.** Four-point probe conductance and XRF thickness measurements were performed within the measurement region to calculate a conductivity value for the film.

# Supplementary references

Jain, S. R., K. C. Adiga, and V. R. Pai Verneker. 1981. "A New Approach to Thermochemical Calculations of Condensed Fuel-Oxidizer Mixtures." *Combustion and Flame* 40 (January): 71–79.

We gratefully acknowledge the efforts of the contributors to the many open-source software packages used in this work:

- python (https://www.python.org/)
- NumPy (https://numpy.org/)
- pandas (https://pandas.pydata.org/)
- Matplotlib (https://matplotlib.org/)
- SciPy (https://www.scipy.org/)
- scikit-learn (https://scikit-learn.org/)
- Keras (https://keras.io/)
- TensorFlow (https://www.tensorflow.org/)
- Plotly (https://plotly.com/)
- LMFIT (https://lmfit.github.io/lmfit-py/)
- HyperSpy (https://hyperspy.org/)
- luigi (https://github.com/spotify/luigi)
- Ax (https://ax.dev/)
- BoTorch (https://botorch.org/)
- PyTorch (https://pytorch.org/)
- Flask (https://flask.palletsprojects.com/)
- OBS Studio (https://obsproject.com/)
- Git (https://git-scm.com/)